\documentclass[12pt,a4paper]{article}
\usepackage{amsmath,amsfonts,amsthm,graphicx,lscape}
\numberwithin{equation}{section}

\def\beqa{\begin{eqnarray}}
\def\enqa{\end{eqnarray}}
\def\beq{\begin{equation}}
\def\enq{\end{equation}}

\begin{document}
\title{
 Nonlinear Integral Equations and high temperature expansion for the $U_{q}(\widehat{sl}(r+1|s+1))$ 
 Perk-Schultz Model
}
\author{Zengo Tsuboi 
\footnote{
E-mail address: 
zengo\_tsuboi@pref.okayama.jp
}
\\
{\it Okayama Institute for Quantum Physics,}
 \\
{\it Kyoyama 1-9-1, Okayama 700-0015, Japan}
}
\date{}
\maketitle
\begin{abstract}
We propose a system of nonlinear integral equations (NLIE) 
which gives the free energy of the 
$U_{q}(\widehat{sl}(r+1|s+1))$ Perk-Schultz model. 
In contrast with traditional thermodynamic Bethe ansatz 
equations, our NLIE contain only $r+s+1$ unknown functions. 
In deriving the NLIE, the 
 quantum (supersymmetric) Jacobi-Trudi and Giambelli formula 
and a duality for an auxiliary function play important roles.  
By using our NLIE, we also calculate the 
high temperature expansion of the free energy. 
General formulae of the coefficients 
with respect to arbitrarily rank $r+s+1$, chemical potentials 
$\{\mu_{a}\}$ and $q$  
have been written down in terms of characters up to the order of 5. 
In particular for 
 specific values of the parameters, 
we have calculated the high temperature expansion 
 of the specific heat up to the order of 40. 
\end{abstract}
{\it MSC:} 82B23; 45G15; 82B20; 17B80 \\
{\it PACS2003:} 02.30.Rz; 02.30.Ik; 02.20.Uw; 05.70.-a \\
{\it Key words:}
nonlinear integral equation; 
Perk-Schultz model; 
quantum Jacobi-Trudi and Giambelli formula; 
quantum transfer matrix; 
thermodynamic Bethe ansatz; 
$T$-system \\
{\bf OIQP-05-12} 
\section{Introduction} 
In statistical physics, 
to calculate the free energy of solvable lattice models
 for finite temperature is one of the 
 important problems. 
 For this purpose, thermodynamic Bethe ansatz (TBA) 
equations have been often used \cite{Ta99}. In general, 
the TBA equations are an infinite number of coupled nonlinear 
integral equations (NLIE) with an infinite number unknown functions. 
Then it is desirable to reduce TBA equations 
to a finite number of coupled NLIE with a finite number of 
unknown functions. 

Destri, de Vega \cite{DD92} and Kl\"umper \cite{K93,K92}
 proposed NLIE with two (or one \footnote{ if an integral contour 
 with a closed loop is adopted}) 
unknown functions for the $XXZ$ (or $XYZ$) model. 
To generalize their NLIE to models 
 whose underlying algebras have arbitrary rank seems to be 
a difficult problem as we need considerable trial and errors 
to find auxiliary functions which are needed to derive the NLIE.  
Then there are NLIE of abovementioned type
 for models whose underlying algebras have 
at most rank 3 (for example, \cite{KWZ97,FK99,D05}). 

Several years ago, 
Takahashi discovered \cite{Ta01} an another NLIE for the 
$XXZ$ model in simplifying TBA equations. 
Later, the same NLIE was rederived \cite{TSK01} from fusion relations 
($T$-system) \cite{KR87} 
among quantum transfer matrices (QTM) \cite{S85}. 
In addition, it was also rederived \cite{KW02} for the $XXX$ model  
from a fugacity expansion formula. 

In view of these situations, we have derived NLIE of Takahashi type for 
the $osp(1|2s)$ model \cite{T02}, the $sl(r+1)$ model \cite{T03},
the higher spin Heisenberg model \cite{T04}, 
the $U_{q}(\widehat{sl}(r+1))$ Perk-Schultz model \cite{TT05}. 
In these cases, 
the number of unknown functions and NLIE coincide with the rank of the 
underlying algebras. In this paper, we will further derive NLIE with a finite 
number of unknown functions 
for the $U_{q}(\widehat{sl}(r+1|s+1))$ Perk-Schultz model \cite{PS81,Sc83}, 
which is a multicomponent generalization of the 6-vertex model and 
one of the fundamental solvable lattice models in statistical mechanics. 
For example, a special case of this model is related to the 
supersymmetric $t-J$ model, which is important in 
strongly correlated electron systems. 

In section 2, we introduce the $U_{q}(\widehat{sl}(r+1|s+1))$ Perk-Schultz model, 
and define the QTM for it. 
As a summation over tableaux labeled by $a \times m$ Young (super) diagram, we 
introduce an auxiliary function (\ref{DVF}) \cite{T97,T98}
 which includes an eigenvalue formula (\ref{QTM-eigen}) of the QTM 
as a special case. 
We also introduce a system of functional relations ($T$-system) which is satisfied by
 this auxiliary function.

In section 3, we derive two kind of NLIE which contain only $r+s+1$ unknown functions. 
The first ones (\ref{nlie-general}), (\ref{nlie-generalb})  
 reduce to the NLIE for the $U_{q}(\widehat{sl}(r+1))$ Perk-Schultz model 
in \cite{TT05} if $s=-1$. 
However our new NLIE are not straightforward generalization of the ones in 
our previous paper \cite{TT05}. 
In fact for $r,s \ge 0$ case, 
a straightforward generation of our previous NLIE 
 becomes a system of an infinite number of coupled NLIE which contains an 
infinite number of unknown functions 
(see (\ref{nlie4})). 
To overcome this difficulty, we will use the 
quantum (supersymmetric) Jacobi-Trudi and Giambelli  formula 
(\ref{jacobi-trudi}) and 
a duality (\ref{dual}) for the auxiliary function, 
from which a closed set of NLIE can be derived. 
We will also propose another NLIE (\ref{nlie-xi=-1})  and  (\ref{nlie-xi=-1b}) 
in the latter part of the 
section 3, which have never been considered before 
even for the $U_{q}(\widehat{sl}(2))$ case. 
In deriving the NLIE, we assume that $q$ is generic. 
However we expect that our results can be also analytically continued to 
the case where $q$ is a root of unity. 

In section 4, we calculate the high temperature expansion of the 
free energy based on our NLIE.
 In particular, we can derive coefficients (\ref{coe1})-(\ref{coe5}) 
up to the order of 5 for the arbitrary rank $r+s+1$. 
The point is that if we fix the degree of the high temperature expansion, 
we can write down a general formula of the coefficients. 
On the other hand, if we specialize parameters, we can 
derive the coefficients for much higher orders. For example 
for $(r,s)=(2,-1),(-1,2)$, $q=1$, $\mu_{a}=0$ case, 
 coefficients of the high temperature expansion of the specific heat 
 up to the order of 40 are presented in appendix. 
 It will be difficult to derive the coefficients of such a 
high order by other method. 
 
Section 5 is devoted to concluding remarks. 
\section{The Perk-Schultz model and the quantum transfer matrix method} 
In this section, we  will introduce the $U_{q}(\widehat{sl}(r+1|s+1))$  
Perk-Schultz model
\footnote{$U_{q}(\widehat{sl}(r+1|s+1))$ is a quantum affine superalgebra, 
which characterizes the $R$-matrix of this model. 
See for example, \cite{Y99}. 
We assume $\eta \in {\mathbb R}$ ($q=e^{\eta}$). 
A rational limit ($q \to 1$) of the Perk-Schultz model is
 the Uimin-Sutherland model \cite{U70,S75}.}
 \cite{PS81,Sc83} and 
the quantum transfer matrix (QTM) method 
 \cite{S85,SI87,K87,SAW90,K92,K93} 
 for it. 
The QTM method was applied to the Perk-Schultz model 
 in ref. \cite{KWZ97} 
(see also, ref. \cite{JKS97,JKS98,FK99}). 

Let us introduce three sets $B=\{1,2,\dots,r+s+2\}=B_{+}\cup B_{-}$, 
where $B_{+} \cap B_{-}=\phi $, $|B_{+}|=r+1$ and $|B_{-}|=s+1$
 ($r,s \in {\mathbb Z}_{\ge -1}$).
We define a grading parameter $p(a)$ ($a \in B$) such that 
$p(a)=0$ for $a \in B_{+}$ and 
$p(a)=1$ for $a \in B_{-}$.  
The $R$-matrix of the $U_{q}(\widehat{sl}(r+1|s+1))$ Perk-Schultz model \cite{PS81} 
is given as 
\begin{eqnarray}
R(v)=
\sum_{a_{1},a_{2},b_{1},b_{2}\in B}
R^{a_{1},b_{1}}_{a_{2},b_{2}}(v) 
E^{a_{1},a_{2}}\otimes E^{b_{1},b_{2}},
\end{eqnarray}
where $E^{a,b}$ is a $r+s+2$ by $r+s+2$ matrix 
whose $(i,j)$ element is given as 
$(E^{a,b})_{i,j}=\delta_{ai}\delta_{bj}$; 
$R^{a_{1},b_{1}}_{a_{2},b_{2}}(v)$ is defined as 
\begin{eqnarray}
&& R^{a,a}_{a,a}(v)=[(-1)^{p(a)}v+1]_{q}, \\
&& R^{a,b}_{a,b}(v)=(-1)^{p(a)p(b)}[v]_{q} \quad (a \ne b), \\
&& R^{b,a}_{a,b}(v)=q^{\mathrm{sign}(a-b)v}
\quad (a \ne b), \label{R-mat}
\end{eqnarray}
where $v \in \mathbb{C}$ is the spectral parameter;
$a,b \in B$; 
 $[v]_{q}=(q^{v}-q^{-v})/(q-q^{-1})$; 
$q=e^{\eta}$. 
Note that this $R$-matrix reduces to the one for the well known 6-vertex model 
if $(r,s)=(1,-1)$.

Let $L$ be a positive integer (the number of lattice sites). 
The row-to-row transfer matrix on $({\mathbb C}^{r+s+2})^{\otimes L}$ 
is defined as
\footnote{The lower index $i,j$ of $R_{ij}(v)$ is used as follows:  
for example, $E^{a,b}_{k}$ 
is defined on $({\mathbb C}^{r+s+2})^{\otimes (L+1)}$:  
$E^{a,b}_{k}=I^{\otimes k}\otimes E^{a,b}\otimes I^{\otimes (L-k)}$, 
where $I$ is $r+s+2$ by $r+s+2$ identity matrix; 
$k=0,1,\dots, L$. 
Then 
$R_{ij}(v)$ is defined as 
$
R_{ij}(v)=\sum_{a_{1},a_{2},b_{1},b_{2}} 
R^{a_{1},b_{1}}_{a_{2},b_{2}}(v) 
E^{a_{1},a_{2}}_{i} E^{b_{1},b_{2}}_{j}
$. The trace ${\mathrm tr}_{0}$ is 
taken over the auxiliary space indexed by $0$.} 
\begin{eqnarray}
t(v)={\mathrm tr}_{0}(R_{0L}(v)
 \cdots R_{02}(v)R_{01}(v)).
\label{rtr}
\end{eqnarray}
The main part of the Hamiltonian is proportional to 
the logarithmic derivative of the row-to-row transfer matrix (\ref{rtr}): 
\begin{eqnarray}
&& \hspace{-20pt} 
H_{body}=\frac{J\sinh \eta}{\eta}\frac{d}{dv}\log t(v) |_{v=0} 
= J\sum_{j=1}^{L}\biggl\{
 \cosh \eta \sum_{a \in B} (-1)^{p(a)}E^{a,a}_{j}E^{a,a}_{j+1} +
\nonumber \\ && 
  \sum_{
  {\scriptsize \begin{array}{c}
  a, b \in B \\
  a\ne b 
  \end{array}}
  }
 \left( {\rm sign}(a-b) \sinh \eta 
  E^{a,a}_{j}E^{b,b}_{j+1} +
 (-1)^{p(a)p(b)}E^{b,a}_{j}E^{a,b}_{j+1}
 \right)
\biggl\},  \label{ham0}
\end{eqnarray}
where we adopt the  periodic boundary condition 
$E^{a,b}_{L+1}=E^{a,b}_{1}$. 
Without breaking the integrability, we can also add the chemical 
potential term
\begin{eqnarray}
H_{ch}=-\sum_{j=1}^{L}\sum_{a \in B}\mu_{a}E^{a,a}_{j} \label{hamch}
\end{eqnarray}
 to $H_{body}$. Then the total Hamiltonian is $H=H_{body}+H_{ch}$. 
 
To treat the model at finite temperature $T$,  
we introduce the so-called quantum transfer matrix (QTM)\cite{S85}: 
\begin{eqnarray}
&& \hspace{-30pt} t_{\mathrm{QTM}}(v)=\sum_{\{\alpha_{k}\},\{\beta_{k}\}}
t_{\mathrm{QTM}}(v)
^{\{\beta_{1},\dots, \beta_{N} \}}
_{\{\alpha_{1},\dots,\alpha_{N} \}}
E^{\beta_{1}\alpha_{1}}_{1}
E^{\beta_{2}\alpha_{2}}_{2}
\cdots 
E^{\beta_{N}\alpha_{N}}_{N}, \label{QTM} \\
&& \hspace{-46pt}
t_{\mathrm{QTM}}(v)^{\{\beta_{1},\dots, \beta_{N} \}}
_{\{\alpha_{1},\dots,\alpha_{N} \}}=
\sum_{\{\nu_{k}\}}e^{\frac{\mu_{\nu_{1}}}{T}}
\prod_{k=1}^{\frac{N}{2}}
 R^{\beta_{2k},\nu_{2k+1}}_{\alpha_{2k},\nu_{2k}}(u+iv)
 \widetilde{R}^{\beta_{2k-1},\nu_{2k}}_{\alpha_{2k-1},\nu_{2k-1}}(u-iv),
 \nonumber 
\end{eqnarray}
where $N \in 2{\mathbb Z}_{\ge 1} $ is the Trotter number; 
$\nu_{N+1}=\nu_{1}$; $\nu_{k},\alpha_{k},\beta_{k}
 \in B$; $u=-\frac{J \sinh \eta }{\eta N T}$; 
$\widetilde{R}^{a_{1},b_{1}}_{a_{2},b_{2}}(v)=
R^{b_{1},a_{2}}_{b_{2},a_{1}}(v)$ is the \symbol{"60}$90^{\circ}$ rotation' of $R(v)$. 
 We can express \cite{S85} the free energy per site 
in terms of only the largest eigenvalue $\Lambda_{1}$ of 
the QTM (\ref{QTM}) at $v=0$:
\begin{eqnarray}
f=
-T\lim_{N\to \infty}\log \Lambda_{1},
\label{free-en-qtm}
\end{eqnarray} 
where the Boltzmann constant is set to $1$. 

Due to the Yang-Baxter equation, the QTM (\ref{QTM}) forms 
commuting family for any $v$. 
Thus it can be diagonalized by the 
Bethe ansatz.  
The eigenvalue formula
\footnote{To be precise, 
this formula is a conjecture 
for general parameters $r,s,q,\mu_{a},N$. 
In \cite{KWZ97}, the 
algebraic Bethe ansatz for a one particle state was 
executed for the QTM of the $U_{q}(\hat{sl}(r+1|s+1))$ Perk-Schultz model. 
As for the $U_{q}(\hat{sl}(2))$ case, a proof of this formula by 
 the algebraic Bethe ansatz is similar to the 
row-to-row transfer matrix case (cf. \cite{GKS04}). 
This formula  has a quite natural form (dressed vacuum form) 
from a point of view of the analytic Bethe ansatz \cite{R83,KS95}. 
An eigenvalue formula of the row to row transfer matrix (\ref{rtr}) 
was derived in \cite{BVV82,Sc83}. It has essentially same form as
 (\ref{QTM-eigen}) except for a part which is related to 
 the vacuum eigenvalue. 
There is also support by numerical calculations for small 
$r,s$.}
 of the QTM (\ref{QTM}) will be (cf. \cite{KWZ97,FK99}) 
\begin{eqnarray}
T^{(1)}_{1}(v)=\sum_{a\in B}z(a;v), 
 \label{QTM-eigen}
\end{eqnarray}
where 
\begin{eqnarray}
&& z(a;v)=\psi_{a}(v) \xi_{a}
 \nonumber \\ 
&& \times 
\frac{Q_{a-1}(v-\frac{i\sum_{j=1}^{a-1}(-1)^{p(j)}}{2}-i(-1)^{p(a)})
Q_{a}(v-\frac{i\sum_{j=1}^{a}(-1)^{p(j)}}{2}+i (-1)^{p(a)})}
{Q_{a-1}(v-\frac{i\sum_{j=1}^{a-1}(-1)^{p(j)}}{2})
Q_{a}(v-\frac{i\sum_{j=1}^{a}(-1)^{p(j)}}{2})},  \nonumber \\
&& Q_{a}(v)=\prod_{k=1}^{M_{a}}\sin \eta(v-v_{k}^{(a)}),
   \\
&& \psi_{a}(v)=e^{\frac{\mu_{a}}{T}}
 \phi_{-}(v-i(-1)^{p(1)}\delta_{a,1})
\phi_{+}(v+i(-1)^{p(r+s+2)}\delta_{a,r+s+2}),
 \nonumber \\
&& \hspace{20pt}
\phi_{\pm}(v)=\left(
\frac{\sin \eta (v\pm iu)}{\sinh \eta }\right)^{\frac{N}{2}},
\nonumber 
\end{eqnarray}
where $M_{a}\in {\mathbb Z}_{\ge 0}$; $Q_{0}(v)=Q_{r+s+2}(v)=1$. 
$\xi_{a} \in \{-1,1\}$ is a parameter which depends on the grading 
parameter $\{p(b)\}_{b \in B}$. 
$\{v^{(a)}_{k}\}$ is a root of the Bethe ansatz equation 
(BAE)
\begin{eqnarray}
&& \hspace{-20pt} 
\frac{\psi_{a}(v^{(a)}_{k}+\frac{i}{2}\sum_{j=1}^{a}(-1)^{p(j)})}
     {\psi_{a+1}(v^{(a)}_{k}+\frac{i}{2}\sum_{j=1}^{a}(-1)^{p(j)})} \label{BAE} \\
&& =
-\varepsilon_{a}
\frac{Q_{a-1}(v^{(a)}_{k}+\frac{i(-1)^{p(a)}}{2})Q_{a}(v^{(a)}_{k}-i(-1)^{p(a+1)})
      Q_{a+1}(v^{(a)}_{k}+\frac{i(-1)^{p(a+1)}}{2})}
      {Q_{a-1}(v^{(a)}_{k}-\frac{i(-1)^{p(a)}}{2})Q_{a}(v^{(a)}_{k}+i(-1)^{p(a)})
      Q_{a+1}(v^{(a)}_{k}-\frac{i(-1)^{p(a+1)}}{2})}
     \nonumber \\
&& \hspace{40pt} \mbox{for} \quad k\in \{1,2, \dots, M_{a}\} \quad 
\mbox{and} \quad a\in \{1,2,\dots, r+s+1 \}, \nonumber
\end{eqnarray}
where $\varepsilon_{a}=\frac{\xi_{a+1}}{\xi_{a}} \in \{-1,1 \} $.
From now on, we assume the relation $p(1)=p(r+s+2)$ on 
the grading parameter.  
In this case, the eigenvalue formula (\ref{QTM-eigen}) 
of the QTM has good analyticity to derive the NLIE. 
We expect that this assumption does not spoil generality 
as the free energy will be independent of the order of the 
grading parameters. 

Let us define 
an auxiliary function \cite{T97,T98} (see also \cite{T98-2}): 
\begin{eqnarray}
T_{m}^{(a)}(v)=\sum_{\{d_{j,k}\}} \prod_{j=1}^{a}\prod_{k=1}^{m}
z(d_{j,k};v-\frac{i}{2}(a-m-2j+2k)),
\label{DVF}
\end{eqnarray}
where $m,a \in \mathbb{Z}_{\ge 1}$, and the summation is taken over 
$d_{j,k}\in B$ ($ 1 < 2 < \cdots < r+s+2$) 
such that
\begin{eqnarray}
&& d_{j,k} \le d_{j+1,k} \quad {\rm and} \quad d_{j,k} \le d_{j,k+1} \label{rule1} \\ 
&& d_{j,k} < d_{j,k+1} \quad {\rm if} \quad 
 d_{j,k} \in B_{-} \quad {\rm or} \quad d_{j,k+1} \in B_{-} 
 \label{rule2} \\  
&& d_{j,k} < d_{j+1,k} \quad {\rm if} \quad  d_{j,k} \in B_{+} 
\quad {\rm or} \quad d_{j+1,k} \in B_{+}. \label{rule3}
\end{eqnarray}
This function contains 
$T_{1}^{(1)}(v)$ (\ref{QTM-eigen}) as a special case 
$(a,m)=(1,1)$. 
(\ref{DVF}) can be interpreted as a 
summation over a Young (super) tableaux labeled by 
$a \times m$ Young (super) diagram. 
It is related to a system of eigenvalue formulae of the 
QTM for fusion models \cite{KRS81}. 
Note that the condition (\ref{rule2}) is void if $s=-1$, then 
(\ref{DVF}) reduces to the Bazhanov-Reshetikhin formula \cite{BR90}. 

For $a,m \in {\mathbb Z}_{\ge 1}$, we 
 will normalize (\ref{DVF}) as 
 $ \widetilde{T}^{(a)}_{m}(v)=
 T^{(a)}_{m}(v)/{\mathcal N}^{(a)}_{m}(v)$, 
 where 
\begin{eqnarray}
\hspace{-30pt} && {\mathcal N}^{(a)}_{m}(v)=
  \frac{\phi_{-}(v- \frac{a+m}{2} \xi i)
\phi_{+}(v+ \frac{a+m}{2}\xi i)}{
  \phi_{-}(v-\frac{a-m}{2}i)\phi_{+}(v+\frac{a-m}{2}i)}
  \nonumber \\ 
\hspace{-30pt}  && \hspace{20pt} \times
  \prod_{j=1}^{a}\prod_{k=1}^{m}
  \phi_{-}(v-\frac{a-m-2j+2k}{2}i)\phi_{+}(v-\frac{a-m-2j+2k}{2}i).
  \label{normal}
\end{eqnarray}
Here we introduce a parameter $\xi \in \{-1,1 \}$. 
$T^{(a)}_{m}(v)$ has no pole on $v$ due to the BAE (\ref{BAE}). 
In contrast, $\widetilde{T}^{(a)}_{m}(v)$ has 
poles at $v=\pm (\frac{m+a}{2}\xi i +iu)+\frac{n \pi}{\eta}$  
($n \in {\mathbb Z}$) for 
$(a,m) \in {\mathbb Z}_{\ge 1} \times \{1,2,\dots,s+1 \} \cup 
 \{1,2,\dots,r+1 \}\times {\mathbb Z}_{\ge 1}$.  

One can show that 
$\widetilde{T}^{(a)}_{m}(v)$ satisfies the 
so called $T$-system for $U_{q}(\widehat{sl}(r+1|s+1))$ \cite{T97,T98} 
(see also \cite{JKS98} for a derivation of TBA equations from the 
$T$-system). 
For  $m,a \in {\mathbb Z}_{\ge 1}$,
\begin{eqnarray}
&& \hspace{-10pt} 
\widetilde{T}^{(a)}_{m}(v-\frac{i}{2})\widetilde{T}^{(a)}_{m}(v+\frac{i}{2})=
\widetilde{T}^{(a)}_{m-1}(v)\widetilde{T}^{(a)}_{m+1}(v)+
\widetilde{T}^{(a-1)}_{m}(v)\widetilde{T}^{(a+1)}_{m}(v)\label{T-sys} \\ 
&& \hspace{-10pt} \mbox{for} \quad 
a \in \{1,2,\dots, r\} \quad \mbox{or} \quad m \in \{1,2,\dots, s\}
 \quad \mbox{or}\quad  (a,m)=(r+1,s+1), \nonumber \\
&& \hspace{-10pt}
\widetilde{T}^{(r+1)}_{m}(v-\frac{i}{2})\widetilde{T}^{(r+1)}_{m}(v+\frac{i}{2})=
\widetilde{T}^{(r+1)}_{m-1}(v)\widetilde{T}^{(r+1)}_{m+1}(v) 
\quad \mbox{for} \quad m \in {\mathbb Z}_{\ge s+2}, \label{T-sys-m} \\
&& \hspace{-10pt}
\widetilde{T}^{(a)}_{s+1}(v-\frac{i}{2})\widetilde{T}^{(a)}_{s+1}(v+\frac{i}{2})=
\widetilde{T}^{(a-1)}_{s+1}(v)\widetilde{T}^{(a+1)}_{s+1}(v) 
\quad \mbox{for} \quad a \in {\mathbb Z}_{\ge r+2}, \label{T-sys-a}
\end{eqnarray}
where 
\begin{eqnarray}
&& \hspace{-35pt} 
 \widetilde{T}^{(a)}_{0}(v)=\frac{\phi_{-}(v-\frac{a}{2}i)\phi_{+}(v+\frac{a}{2}i)}
  {\phi_{-}(v-\frac{a}{2}\xi i)\phi_{+}(v+\frac{a}{2} \xi i)}
\quad {\rm for} \quad a \in {\mathbb Z}_{\ge 1},\label{a0} \\
&& \hspace{-35pt}
\widetilde{T}^{(0)}_{m}(v)=
 \frac{\phi_{-}(v+\frac{m}{2}i)\phi_{+}(v-\frac{m}{2}i)}
  {\phi_{-}(v-\frac{m}{2} \xi i)\phi_{+}(v+\frac{m}{2} \xi i)}
\quad {\rm for} \quad m \in {\mathbb Z}_{\ge 1}. \label{0m} 
\end{eqnarray}
There is a duality relation for the auxiliary function.
\begin{eqnarray} 
&& \hspace{-35pt}
\widetilde{T}^{(r+1)}_{a+s}(v)=
\zeta^{a-1} 
\widetilde{T}^{(r+a)}_{s+1}(v) \quad  {\rm for} \quad a \in Z_{\ge 1} ,
\label{dual}
\end{eqnarray} 
where 
$\zeta = \frac{\prod_{a \in B_{+}} \xi_{a}
 e^{\frac{\mu_{a}}{T}}}{\prod_{b \in B_{-}}\xi_{b}e^{\frac{\mu_{b}}{T}}}$. 
(\ref{a0}) (resp. (\ref{0m})) becomes $1$ if $\xi=1$ (resp. $\xi=-1$). 
Note that there is no upper bound for the index $a$ of $\widetilde{T}^{(a)}_{m}(v)$ 
for $m \in \{1,2,\dots, s+1 \}$ if $s \in {\mathbb Z}_{\ge 0}$.
For $s=-1$, this $T$-system reduces the one for $U_{q}(\widehat{sl}(r+1))$
 \cite{KNS94} (see also \cite{KR87}). 
In this case, (\ref{dual}) reduces to 
$\widetilde{T}^{(r+1)}_{a-1}(v)=\zeta^{a-1}=
e^{\frac{(a-1)(\mu_{1}+\mu_{2}+\cdots +\mu_{r+1})}{T}}$ 
 if $ \xi =1 $ (see eq. (2.21) in \cite{TT05}).
From the relations (\ref{T-sys-m}), (\ref{T-sys-a}), (\ref{dual}) and 
(\ref{T-sys}) for $(a,m)=(r+1,s+1)$, one can derive the following relation 
 for $a \in {\mathbb Z}_{\ge 2}$: 
\begin{eqnarray}
&& \hspace{-20pt} \widetilde{T}^{(r+1)}_{s+a}(v) =
\zeta^{a-1}
\widetilde{T}^{(r+a)}_{s+1}(v) \nonumber \\
&& =
 \frac{
\zeta^{a-1} 
\prod_{j=1}^{a} \widetilde{T}^{(r+1)}_{s+1}(v+\frac{a-2j+1}{2}i) }
 {\prod_{j=2}^{a} \bigl( 
 \zeta
\widetilde{T}^{(r+1)}_{s}(v+\frac{a-2j+2}{2}i)+
 \widetilde{T}^{(r)}_{s+1}(v+\frac{a-2j+2}{2}i) \bigr)}  .
 \nonumber \\
  \label{sol}
\end{eqnarray}
$\widetilde{T}^{(a)}_{m}(v)$ can also be written in terms of a determinant 
(the quantum (supersymmetric) Jacobi-Trudi and Giambelli formula \cite{T97,T98} 
(for $s=-1$ case, \cite{BR90}; 
for $U_{q}(B_{r}^{(1)})$ case, \cite{KOS95}))
\begin{eqnarray}
\widetilde{T}^{(a)}_{m}(v)&=&
 W^{(a)}_{m}(v)\det _{1\le j,k \le m}
\left(\widetilde{T}^{(a+j-k)}_{1}
\left(
v-\frac{j+k-m-1}{2}i
\right) 
\right) \label{jacobi-trudi} \\
&=&  Z^{(a)}_{m}(v) \det _{1\le j,k \le a}
\left(\widetilde{T}^{(1)}_{m+j-k}
\left(
v-\frac{a-j-k+1}{2}i
\right) 
\right), \label{jacobi-trudi2}
\end{eqnarray}
where $\widetilde{T}^{(a)}_{1}(v)=0$ for $a <0$ and 
$\widetilde{T}^{(1)}_{m}(v)=0$ for $m <0$. 
$ W^{(a)}_{m}(v)$ and $ Z^{(a)}_{m}(v)$ are normalization functions: 
\begin{eqnarray}
&& W^{(a)}_{m}(v)=\frac{1}{\prod_{j=1}^{m-1}\widetilde{T}^{(a)}_{0}(v+\frac{m-2j}{2}i)}, \\
&&  Z^{(a)}_{m}(v)= \frac{1}{\prod_{j=1}^{a-1}\widetilde{T}^{(0)}_{m}(v-\frac{a-2j}{2}i)},
\end{eqnarray} 
where $\prod_{j=1}^{0}(\cdots )=1$. 
Substituting (\ref{jacobi-trudi}) into (\ref{dual}), we obtain an equation
\begin{eqnarray}
&& W^{(r+1)}_{a+s}(v) \det _{1\le j,k \le a+s}
\left(\widetilde{T}^{(r+1+j-k)}_{1}
\left(
v-\frac{j+k-a-s-1}{2}i
\right) 
\right)  \nonumber \\
&&=
\zeta^{a-1}
W^{(r+a)}_{s+1}(v)
\det _{1\le j,k \le s+1}
\left(\widetilde{T}^{(r+a+j-k)}_{1}
\left(
v-\frac{j+k-s-2}{2}i
\right) 
\right) 
\nonumber \\
&& \hspace{180pt} \mbox{for} \quad 
a \in {\mathbb Z}_{\ge 1}. \label{det-eq}
\end{eqnarray}
Expanding partially (\ref{det-eq}) on both side, 
we obtain
\begin{eqnarray}
&& \widetilde{T}^{(a+r+s)}_{1}(v)=
\frac{
\widetilde{A}_{1}(v)-
\zeta^{a-1}
\frac{W^{(r+a)}_{s+1}(v)}{W^{(r+1)}_{a+s}(v)}
\widetilde{A}_{2}(v)
}
{(-1)^{a+s}\widetilde{A}_{3}(v)+(-1)^{s}
\zeta^{a-1} 
\frac{W^{(r+a)}_{s+1}(v)}{W^{(r+1)}_{a+s}(v)}
 \widetilde{A}_{4}(v)} 
 \nonumber \\ 
&& \hspace{160pt} \mbox{for} \quad 
a \in {\mathbb Z}_{\ge 2}, 
\label{a+r+s}
\end{eqnarray}
where 
\begin{eqnarray}
&& \widetilde{A}_{1}(v)=\det _{1\le j,k \le a+s}
\left(\widetilde{f}_{j,k}
\left(
v-\frac{j+k-a-s-1}{2}i
\right) 
\right) \\
&& \quad \widetilde{f}_{j,k}(v)=\widetilde{T}^{(r+1+j-k)}_{1}(v) 
 \quad \mbox{for} \quad (j,k) \ne (a+s,1), 
\quad  \widetilde{f}_{a+s,1}(v)=0,  \nonumber \\
&&\widetilde{A}_{2}(v)=
\det _{1\le j,k \le s+1}
\left(\widetilde{g}_{j,k}
\left(
v-\frac{j+k-s-2}{2}i
\right) 
\right) 
 \\
&& \quad \widetilde{g}_{j,k}(v)=\widetilde{T}^{(r+a+j-k)}_{1}(v) 
 \quad \mbox{for} \quad (j,k) \ne (s+1,1), 
\quad  \widetilde{g}_{s+1,1}(v)=0,  \nonumber \\ 
&& \widetilde{A}_{3}(v)=\det _{1\le j,k \le a+s-1}
\left(\widetilde{T}^{(r+j-k)}_{1}
\left(
v-\frac{j+k-a-s}{2}i
\right) 
\right), \\
&&\widetilde{A}_{4}(v)=
\det _{1\le j,k \le s}
\left(\widetilde{T}^{(r+a+j-k-1)}_{1}
\left(
v-\frac{j+k-s-1}{2}i
\right) 
\right) 
.
\end{eqnarray}
It turns out that $\widetilde{T}^{(a+r+s)}_{1}(v)$ is written in 
terms of $\{\widetilde{T}^{(d)}_{1}(v)\}$ where $ \max (0,r-s+2-a) \le d \le a+r+s-1$. 
Then  $ \widetilde{T}^{(a)}_{1}(v) $ for $a \in {\mathbb Z}_{\ge r+s+2}$  
can be expressed in 
terms of $\{\widetilde{T}^{(d)}_{1}(v)\}$ where $ 0 \le d \le r+s+1$.
Similarly, we can derive the 
following relation from (\ref{dual}) and (\ref{jacobi-trudi2}).
\begin{eqnarray}
&& \widetilde{T}^{(1)}_{a+r+s}(v)=
\frac{
\zeta^{a-1}
\frac{Z^{(r+a)}_{s+1}(v)}{Z^{(r+1)}_{a+s}(v)}
\widetilde{A}_{5}(v)-
\widetilde{A}_{6}(v)
}
{(-1)^{a+r}
\zeta^{a-1}
\frac{Z^{(r+a)}_{s+1}(v)}{Z^{(r+1)}_{a+s}(v)}
\widetilde{A}_{7}(v)+(-1)^{r}  
\widetilde{A}_{8}(v)} 
\nonumber \\
&& \hspace{140pt} \mbox{for} \quad 
a \in {\mathbb Z}_{\ge 2}, 
\label{a+r+s-b}
\end{eqnarray}
where 
\begin{eqnarray}
&& \widetilde{A}_{5}(v)=\det _{1\le j,k \le a+r}
\left(\widetilde{h}_{j,k}
\left(
v-\frac{a+r+1-j-k}{2}i
\right) 
\right) \\
&& \quad \widetilde{h}_{j,k}(v)=\widetilde{T}^{(1)}_{s+1+j-k}(v) 
 \quad \mbox{for} \quad (j,k) \ne (a+r,1), 
\quad  \widetilde{h}_{a+r,1}(v)=0,  \nonumber \\
&&\widetilde{A}_{6}(v)=
\det _{1\le j,k \le r+1}
\left(\widetilde{b}_{j,k}
\left(
v-\frac{r+2-j-k}{2}i
\right) 
\right) 
 \\
&& \quad \widetilde{b}_{j,k}(v)=\widetilde{T}^{(1)}_{a+s+j-k}(v) 
 \quad \mbox{for} \quad (j,k) \ne (r+1,1), 
\quad  \widetilde{b}_{r+1,1}(v)=0,  \nonumber \\ 
&& \widetilde{A}_{7}(v)=\det _{1\le j,k \le a+r-1}
\left(\widetilde{T}^{(1)}_{s+j-k}
\left(
v-\frac{a+r-j-k}{2}i
\right) 
\right), \\
&&\widetilde{A}_{8}(v)=
\det _{1\le j,k \le r}
\left(\widetilde{T}^{(1)}_{a+s-1+j-k}
\left(
v-\frac{r+1-j-k}{2}i
\right) 
\right) 
.
\end{eqnarray}

Let us consider the limit 
\begin{eqnarray}
&& Q^{(a)}_{m}:=\lim_{v \to i \eta^{-1} \infty} \widetilde{T}^{(a)}_{m}(v) 
 =\sum_{\{ d_{j,k}\}}
\prod_{j=1}^{a}\prod_{k=1}^{m} \xi_{d_{j,k}}
\exp \left(\frac{\mu_{d_{j,k}}}{T} \right),
 \label{limit}
\end{eqnarray} 
where the summation is taken over $\{ d_{j,k}\}$ ($d_{j,k} \in B$)
 which obey the rules (\ref{rule1})-(\ref{rule3}).
For example, for $U_{q}(\widehat{sl}(2|1))$ ($B_{+}=\{1,3\}$, $B_{-}=\{2\}$) case, we have, 
\begin{eqnarray}
Q^{(1)}_{1}&=& \xi_{1}e^{\frac{\mu_{1}}{T}}+\xi_{2}e^{\frac{\mu_{2}}{T}}
 +\xi_{3} e^{\frac{\mu_{3}}{T}},
 \label{Q11-sl21} \\
Q^{(a)}_{1}&=&
 \xi_{1} \xi_{2}^{a-1} e^{\frac{\mu_{1}+(a-1)\mu_{2}}{T}}
+\xi_{1} \xi_{2}^{a-2} \xi_{3} e^{\frac{\mu_{1}+(a-2)\mu_{2}+\mu_{3}}{T}}
+\xi_{2}^{a}e^{\frac{a \mu_{2}}{T}}
+\xi_{2}^{a-1} \xi_{3} e^{\frac{(a-1)\mu_{2}+\mu_{3}}{T}} \nonumber \\
&=& \xi_{2}^{a-2}e^{ \frac{(a-2) \mu_{2}}{T}}Q^{(2)}_{1} 
\qquad  \mbox{for} \quad a \in {\mathbb Z}_{\ge 2}. 
 \label{Q-sl21}
\end{eqnarray}
We can also rewrite (\ref{Q-sl21}) as  
\begin{eqnarray}
Q^{(a)}_{1}=\frac{{Q^{(3)}_{1}}^{a-2}}{{Q^{(2)}_{1}}^{a-3}}
=\frac{{Q^{(2)}_{1}}^{a-1}}{(\zeta +Q^{(1)}_{1})^{a-2}}.
\label{Qa1-sl21}
\end{eqnarray}
This quantity (\ref{limit}) corresponds to 
 the character of $a$-th anti-(super)symmetric and 
$m$-th (super)symmetric tensor representation. 
We will use  $Q^{(a)}_{1}$ and $Q^{(1)}_{m}$ later. 

$Q^{(a)}_{m}$ also satisfies the so called $Q$-system, 
which is the $T$-system (\ref{T-sys})-(\ref{dual})
 without the spectral parameter $v$:  
for  $ m,a \in {\mathbb Z}_{\ge 1}$, we have 
\begin{eqnarray}
\hspace{-20pt} && {Q^{(a)}_{m}}^{2}=Q^{(a)}_{m-1}Q^{(a)}_{m+1}+Q^{(a-1)}_{m}Q^{(a+1)}_{m}
\label{Q-sys} \\ 
&&\hspace{10pt} \mbox{for} \quad 
a \in \{1,2,\dots, r\} \quad \mbox{or} \quad m \in \{1,2,\dots, s\}
 \nonumber \\
&& \hspace{130pt} \mbox{or}\quad  (a,m)=(r+1,s+1), \nonumber \\
&&{Q^{(r+1)}_{m}}^{2}=Q^{(r+1)}_{m-1}Q^{(r+1)}_{m+1}
 \quad \mbox{for} \quad m \in {\mathbb Z}_{\ge s+2},\\
&&{Q^{(a)}_{s+1}}^{2} =Q^{(a-1)}_{s+1}Q^{(a+1)}_{s+1} 
\quad \mbox{for} \quad a \in {\mathbb Z}_{\ge r+2}, 
\end{eqnarray}
where 
\begin{eqnarray}
&& Q^{(a)}_{0}=Q^{(0)}_{m}=1
\quad {\rm for} \quad a,m \in {\mathbb Z}_{\ge 1},\nonumber \\
&& Q^{(r+1)}_{a+s}=
\zeta^{a-1}
Q^{(r+a)}_{s+1} \quad  {\rm for} \quad a \in Z_{\ge 1} .
\end{eqnarray} 
The $Q$-system was introduced \cite{K89,KR90} as functional relations among 
characters of finite dimensional representations of 
Yangians (or quantum affine algebras) associated with simple Lie algebras. 
The above system of equations is a superalgebra version of them. 

In closing this section, 
let us comment on the analyticity of the auxiliary function (\ref{DVF}). 
As mentioned before, the free energy (\ref{free-en-qtm}) is given only by the 
largest eigenvalue of the QTM (\ref{QTM}). 
Then we are only interested in a root of the BAE (\ref{BAE}) 
which gives the largest eigenvalue of the QTM. 
Judging from numerical calculations \cite{JKS97,JKS98,T03,TT05}, 
such a root will exist in the sector 
$\frac{N}{2}=M_{1}=\cdots =M_{r+s+1}$ of the BAE, 
and it will form \symbol{"60}one-string' on the complex plane. 
For this root, the zeros of the auxiliary function (\ref{DVF}) will 
 exist near the lines ${\rm Im} v= \pm \frac{a+m}{2}$ 
 at least for $\{\mu_{a}\}=\{0\}$ and small $u$ 
(see, figures in \cite{JKS98,T03,TT05}). 
In this sector, we have 
\begin{eqnarray}
&& \xi_{b}=1 \qquad {\rm for} \qquad b \in B,
\nonumber \\ 
&& \varepsilon_{b}=1 \qquad {\rm for} \qquad b \in B-\{r+s+2 \},
\label{para}
\\
&& \zeta=\exp(\frac{\sum_{a\in B_{+}}\mu_{a}-\sum_{a\in B_{-}}\mu_{a}}{T}).
\nonumber \end{eqnarray} 
From now on, we only consider the largest eigenvalue of the 
QTM, and assume these values (\ref{para}) of the parameters. 
\section{The nonlinear integral equations}
In this section, we will derive NLIE by using formulae in the previous section. 
We will treat two kind of NLIE paying attention to the value of the 
parameter $\xi \in \{-1,1\}$. 
Although the first step of calculations (\ref{mustput})-(\ref{nlie2}) is similar to
 $s=-1$ case \cite{TSK01,T03,TT05}, we will present it for reader's convenience. 
 
Taking note on 
the limit (\ref{limit}) and 
 the fact that $\widetilde{T}^{(a)}_{m}(v)$ has 
poles at $v=\pm (\frac{m+a}{2}\xi i +iu)+ \frac{n \pi}{\eta}$ 
($n \in {\mathbb Z}$) 
for $(a,m) \in \{1,2,\dots, r+1\}\times {\mathbb Z}_{\ge 1} \cup 
{\mathbb Z}_{\ge 1} \times \{1,2,\dots, s+1\}$, 
we can expand ${\widetilde T}^{(a)}_{m}(v)$ as follows.
\begin{eqnarray}
&& {\widetilde T}^{(a)}_{m}(v)=Q^{(a)}_{m}
 \label{mustput} \\ 
&& \hspace{20pt} +
\sum_{n \in {\mathbb Z}} 
\sum_{j=1}^{\frac{N}{2}} 
 \left\{ 
\frac{A^{(a)}_{m,j}}{(v-\frac{a+m}{2}\xi i-iu-\frac{\pi n}{\eta})^{j}}
+
\frac{{\bar A}^{(a)}_{m,j}}{(v+\frac{a+m}{2}\xi i+iu+\frac{\pi n}{\eta})^{j}}
\right\},
\nonumber 
\end{eqnarray}
where the coefficients $A^{(a)}_{m,j}, {\bar A}^{(a)}_{m,j} \in {\mathbb C}$ 
can be expressed as contour integrals:
\begin{eqnarray}
&& A^{(a)}_{m,j}= \oint_{{\tilde C}^{(a)}_{m}} \frac{{\mathrm d} v}{2\pi i}
 \widetilde{T}^{(a)}_{m}(v)(v-\frac{a+m}{2}\xi i-iu)^{j-1},\nonumber \\
&& \overline{A}^{(a)}_{m,j}=
 \oint_{\overline{\tilde C}^{(a)}_{m}} \frac{{\mathrm d} v}{2\pi i}
 \widetilde{T}^{(a)}_{m}(v)(v+\frac{a+m}{2}\xi i+iu)^{j-1}.
 \label{coeff}
\end{eqnarray}
Here the contour ${\tilde C}^{(a)}_{m}$ (resp. $\overline{\tilde C}^{(a)}_{m}$) 
is a counterclockwise closed loop 
which surrounds $v=\frac{a+m}{2}\xi i+iu$ (resp. $v=-\frac{a+m}{2}\xi i-iu$) 
and does not surround $v=-\frac{a+m}{2}\xi i-iu-\frac{\pi n}{\eta},
\frac{a+m}{2}\xi i+iu+\frac{\pi k}{\eta},$ 
(resp. $v=\frac{a+m}{2}\xi i+iu+\frac{\pi n}{\eta}, 
-\frac{a+m}{2}\xi i-iu-\frac{\pi k}{\eta}$), where $n \in {\mathbb Z}, k \in  {\mathbb Z}-\{0\} $.
Using the $T$-system (\ref{T-sys})-(\ref{T-sys-a}), 
we can rewrite (\ref{coeff}) as 
\begin{eqnarray}
&& A^{(a)}_{m,j}= \oint_{{\tilde C}^{(a)}_{m}} \frac{{\mathrm d} v}{2\pi i}
 \bigg\{
 \frac{\widetilde{T}^{(a)}_{m-1}(v-\frac{\xi i}{2})
       \widetilde{T}^{(a)}_{m+1}(v-\frac{\xi i}{2})}
      {\widetilde{T}^{(a)}_{m}(v-\xi i)} \nonumber \\
&& \hspace{80pt} +
 \frac{\widetilde{T}^{(a-1)}_{m}(v-\frac{\xi i}{2})
       \widetilde{T}^{(a+1)}_{m}(v-\frac{\xi i}{2})}
      {\widetilde{T}^{(a)}_{m}(v-\xi i)}
 \bigg\}
 (v-\frac{a+m}{2}\xi i-iu)^{j-1},\nonumber \\
&& \overline{A}^{(a)}_{m,j}=
 \oint_{\overline{\tilde C}^{(a)}_{m}} \frac{{\mathrm d} v}{2\pi i}
\bigg\{
 \frac{\widetilde{T}^{(a)}_{m-1}(v+\frac{\xi i}{2})
       \widetilde{T}^{(a)}_{m+1}(v+\frac{\xi i}{2})}
      {\widetilde{T}^{(a)}_{m}(v+\xi i)} 
\label{coeff2} \\
&& \hspace{80pt} +
 \frac{\widetilde{T}^{(a-1)}_{m}(v+\frac{\xi i}{2})
       \widetilde{T}^{(a+1)}_{m}(v+\frac{\xi i}{2})}
      {\widetilde{T}^{(a)}_{m}(v+\xi i)}
 \bigg\}
 (v+\frac{a+m}{2}\xi i+iu)^{j-1},
 \nonumber 
\end{eqnarray}
where we admit $\widetilde{T}^{(b)}_{n}(v)=0$ if 
$(b,n) \in {\mathbb Z }_{\ge r+2}\times {\mathbb Z}_{\ge s+2}$
 (cf. \cite{DM92,MR92}).
Substituting (\ref{coeff2}) into (\ref{mustput}) and taking the summation
 over $j$, we obtain
\begin{eqnarray}
&& \hspace{-30pt}
\widetilde{T}^{(a)}_{m}(v)=Q^{(a)}_{m} \nonumber \\ 
&& +
\sum_{n \in {\mathbb Z}}
\oint_{{\tilde C}^{(a)}_{m}} \frac{{\mathrm d} y}{2\pi i} 
\frac{1-\left(\frac{y}{v-\frac{a+m}{2} \xi i-iu -\frac{\pi n}{\eta}}\right)^{\frac{N}{2}}}
 {v-y-\frac{a+m}{2} \xi i-iu -\frac{\pi n}{\eta}} 
 \nonumber \\
&&\hspace{20pt} \times  
 \bigg\{ 
 \frac{\widetilde{T}^{(a)}_{m-1}(y+\frac{a+m-1}{2} \xi i+iu) 
 \widetilde{T}^{(a)}_{m+1}(y+\frac{a+m-1}{2} \xi i+iu)}
 {\widetilde{T}^{(a)}_{m}(y+\frac{a+m-2}{2} \xi i+iu)} \nonumber \\
 && \hspace{50pt} +
 \frac{\widetilde{T}^{(a-1)}_{m}(y+\frac{a+m-1}{2} \xi i+iu) 
 \widetilde{T}^{(a+1)}_{m}(y+\frac{a+m-1}{2} \xi i+iu)}
 {\widetilde{T}^{(a)}_{m}(y+\frac{a+m-2}{2} \xi i+iu)}
 \bigg\} \nonumber \\
&& +
\sum_{n \in {\mathbb Z}}
\oint_{\overline{\tilde C}^{(a)}_{m}} \frac{{\mathrm d} y}{2\pi i} 
\frac{1-\left(\frac{y}{v+\frac{a+m}{2} \xi i+iu +\frac{\pi n}{\eta}}\right)^{\frac{N}{2}}}
 {v-y+\frac{a+m}{2} \xi i+iu +\frac{\pi n}{\eta}} 
 \nonumber \\
&&\hspace{20pt} \times  
 \bigg\{ 
 \frac{\widetilde{T}^{(a)}_{m-1}(y-\frac{a+m-1}{2} \xi i-iu) 
 \widetilde{T}^{(a)}_{m+1}(y-\frac{a+m-1}{2} \xi i-iu)}
 {\widetilde{T}^{(a)}_{m}(y-\frac{a+m-2}{2} \xi i-iu)}  
\label{nlie1} \\
 && \hspace{50pt} +
 \frac{\widetilde{T}^{(a-1)}_{m}(y-\frac{a+m-1}{2} \xi i-iu) 
 \widetilde{T}^{(a+1)}_{m}(y-\frac{a+m-1}{2} \xi i-iu)}
 {\widetilde{T}^{(a)}_{m}(y-\frac{a+m-2}{2} \xi i-iu)}
 \bigg\}.
 \nonumber
\end{eqnarray}
Here the contours are shifted as follows: 
 the contour ${\tilde C}^{(a)}_{m}$ (resp. $\overline{\tilde C}^{(a)}_{m}$) 
is a counterclockwise closed loop 
which surrounds $y=0 $ (resp. $y=0$)
and does not surround $y=-(a+m)\xi i-2iu-\frac{\pi n}{\eta},\frac{\pi k}{\eta}$ 
(resp. $y=(a+m)\xi i+2iu+\frac{\pi n}{\eta},\frac{\pi k}{\eta}$), 
where $n \in {\mathbb Z}, k \in {\mathbb Z}-\{0 \}$.
We can neglect the terms $\left(\frac{y}{v \pm \frac{a+m}{2} \xi i \pm iu \pm 
\frac{\pi n}{\eta}}\right)^{\frac{N}{2}}$ in (\ref{nlie1}) since the poles at $y=0$ in 
 the two brackets $\{\cdots \}$ 
are canceled by the zeros from these terms. 
By using the following relation
\begin{eqnarray}
\lim_{m \to \infty}
\sum_{n=-m}^{m}\frac{1}{v-\frac{\pi n}{\eta}}
=\frac{\eta}{\tan \eta v},
\end{eqnarray}
we can take the summation over $n \in {\mathbb Z}$.
\begin{eqnarray}
&& \hspace{-30pt}
\widetilde{T}^{(a)}_{m}(v)=Q^{(a)}_{m} \nonumber \\ 
&& +
\oint_{{\tilde C}^{(a)}_{m}} \frac{{\mathrm d} y}{2\pi i} 
\frac{\eta }
 {\tan \eta (v-y-\frac{a+m}{2} \xi i-iu)} 
 \nonumber \\
&&\hspace{20pt} \times  
 \bigg\{ 
 \frac{\widetilde{T}^{(a)}_{m-1}(y+\frac{a+m-1}{2} \xi i+iu) 
 \widetilde{T}^{(a)}_{m+1}(y+\frac{a+m-1}{2} \xi i+iu)}
 {\widetilde{T}^{(a)}_{m}(y+\frac{a+m-2}{2} \xi i+iu)} \nonumber \\
 && \hspace{50pt} +
 \frac{\widetilde{T}^{(a-1)}_{m}(y+\frac{a+m-1}{2} \xi i+iu) 
 \widetilde{T}^{(a+1)}_{m}(y+\frac{a+m-1}{2} \xi i+iu)}
 {\widetilde{T}^{(a)}_{m}(y+\frac{a+m-2}{2} \xi i+iu)}
 \bigg\} \nonumber \\
&& +
\oint_{\overline{\tilde C}^{(a)}_{m}} \frac{{\mathrm d} y}{2\pi i} 
\frac{\eta }
 {\tan \eta (v-y+\frac{a+m}{2} \xi i+iu)} 
 \nonumber \\
&&\hspace{20pt} \times  
 \bigg\{ 
 \frac{\widetilde{T}^{(a)}_{m-1}(y-\frac{a+m-1}{2} \xi i-iu) 
 \widetilde{T}^{(a)}_{m+1}(y-\frac{a+m-1}{2} \xi i-iu)}
 {\widetilde{T}^{(a)}_{m}(y-\frac{a+m-2}{2} \xi i-iu)}  
\label{nlie2} \\
 && \hspace{50pt} +
 \frac{\widetilde{T}^{(a-1)}_{m}(y-\frac{a+m-1}{2} \xi i-iu) 
 \widetilde{T}^{(a+1)}_{m}(y-\frac{a+m-1}{2} \xi i-iu)}
 {\widetilde{T}^{(a)}_{m}(y-\frac{a+m-2}{2} \xi i-iu)}
 \bigg\},
 \nonumber \\
 && {\rm for} \quad (a,m) \in 
\{1,2,\dots,r+1\} \times {\mathbb Z}_{\ge 1} \cup 
{\mathbb Z}_{\ge 1} \times \{1,2,\dots,s+1\}.
 \nonumber 
\end{eqnarray}
In the next subsection, we will consider specializations 
of this system of NLIE (\ref{nlie2}). 
\subsection{The nonlinear integral equations for $\xi=1$}
Let us consider the NLIE (\ref{nlie2}) for $\xi=1$ and $m=1$. 
Taking note on the fact ${\widetilde T}^{(a)}_{0}(v)=1$ (cf.(\ref{a0})), 
we can drop the first terms in the two brackets $\{\cdots \}$ in (\ref{nlie2}) 
since they have no poles at $y=0$.
Then the NLIE (\ref{nlie2}) reduce to the following NLIE on 
${\mathcal T}^{(a)}_{1}(v)=\lim_{N \to \infty}\widetilde{T}^{(a)}_{1}(v)$ 
 after the Trotter limit $N \to \infty $ with $u=-\frac{J \sinh \eta }{\eta N T}$.
\begin{eqnarray}
{\mathcal T}^{(a)}_{1}(v)=Q^{(a)}_{1} 
&+&
\oint_{C^{(a)}_{1}} \frac{{\mathrm d} y}{2\pi i} 
 \frac{\eta 
 \mathcal{T}^{(a-1)}_{1}(y+\frac{i a}{2}) 
 \mathcal{T}^{(a+1)}_{1}(y+\frac{i a}{2})}
 {\tan \eta (v-y-\frac{i(a+1)}{2})
 \mathcal{T}^{(a)}_{1}(y+\frac{i(a-1)}{2})}
 \nonumber \\
&+&
\oint_{\overline{C}^{(a)}_{1}} \frac{{\mathrm d} y}{2\pi i} 
 \frac{\eta 
 \mathcal{T}^{(a-1)}_{1}(y-\frac{i a}{2}) 
 \mathcal{T}^{(a+1)}_{1}(y-\frac{i a}{2})}
 {\tan \eta (v-y+\frac{i(a+1)}{2})
 \mathcal{T}^{(a)}_{1}(y-\frac{i(a-1)}{2})}
 \nonumber \\ 
 && \hspace{120pt} 
 {\rm for} \quad a \in {\mathbb Z}_{\ge 1},
 \label{nlie4}
\end{eqnarray}
where 
the contour $C^{(a)}_{1}$ (resp. $\overline{C}^{(a)}_{1}$) 
is a counterclockwise closed loop around $y=0$ (resp. $y=0$) 
which satisfies the condition 
$y \ne v-\frac{a+1}{2}i+\frac{\pi n}{\eta}$ 
(resp. $y \ne v+\frac{a+1}{2}i+\frac{\pi n}{\eta}$) and 
does not surround 
$z^{(a)}_{1}-\frac{a-1}{2}i+\frac{\pi n}{\eta}$, 
$-(a+1)i
+\frac{\pi n}{\eta}$, $\frac{\pi k}{\eta}$ 
(resp. 
$z^{(a)}_{1}+\frac{a-1}{2}i+\frac{\pi n}{\eta}$, 
$(a+1)i +\frac{\pi n}{\eta}$, $\frac{\pi k}{\eta}$); 
($n \in \mathbb{Z}$, $k \in \mathbb{Z}-\{0\}$). 
Here we put the zeros of $\mathcal{T}^{(a)}_{1}(v)$ as $\{ z^{(a)}_{1} \} $: 
$\mathcal{T}^{(a)}_{1}(z^{(a)}_{1})=0$. 
$\mathcal{T}^{(0)}_{1}(v)$ is a known function:
\begin{eqnarray} 
\mathcal{T}^{(0)}_{1}(v)=
\lim_{N \to \infty} \widetilde{T}^{(0)}_{1}(v)
=\exp \left(\frac{2J (\sinh \eta)^{2} }
{T(\cosh \eta -\cos (2\eta v))}\right).
\end{eqnarray}
Note that (\ref{nlie4}) are an infinite number of couple NLIE 
if $ s \in {\mathbb Z}_{\ge 0} $.  
This situation is quite different from the $U_{q}(\widehat{sl}(r+1))$
 case \cite{TT05,T03,TSK01}.  
However these NLIE are not independent, then 
we will take the first $r+s+1$ of them ((\ref{nlie4}) for $a \in \{1,2,\dots r+s+1 \}$). 
The NLIE for $a=r+s+1$ contains $\mathcal{T}^{(r+s+2)}_{1}(v)$, then we 
will eliminate this by using the relation (\ref{a+r+s}), 
where $W^{(a)}_{m}(v)=1$ for $\xi=1$.
\begin{eqnarray}
&& {\mathcal T}^{(a)}_{1}(v)=Q^{(a)}_{1} 
+
\oint_{C^{(a)}_{1}} \frac{{\mathrm d} y}{2\pi i} 
 \frac{\eta 
 \mathcal{T}^{(a-1)}_{1}(y+\frac{i a}{2}) 
 \mathcal{T}^{(a+1)}_{1}(y+\frac{i a}{2})}
 {\tan \eta (v-y-\frac{i(a+1)}{2})
 \mathcal{T}^{(a)}_{1}(y+\frac{i(a-1)}{2})}
 \nonumber \\
&& \hspace{40pt} +
\oint_{\overline{C}^{(a)}_{1}} \frac{{\mathrm d} y}{2\pi i} 
 \frac{\eta 
 \mathcal{T}^{(a-1)}_{1}(y-\frac{i a}{2}) 
 \mathcal{T}^{(a+1)}_{1}(y-\frac{i a}{2})}
 {\tan \eta (v-y+\frac{i(a+1)}{2})
 \mathcal{T}^{(a)}_{1}(y-\frac{i(a-1)}{2})}
 \nonumber \\ 
 && \hspace{70pt} 
 {\rm for} \quad a \in \{1,2,\dots r+s \},
 \label{nlie-general} \\
 && {\mathcal T}^{(r+s+1)}_{1}(v)=Q^{(r+s+1)}_{1}
 \nonumber \\ 
&&\hspace{20pt} +
\oint_{C^{(r+s+1)}_{1}} \frac{{\mathrm d} y}{2\pi i} 
 \frac{\eta 
 \mathcal{T}^{(r+s)}_{1}(y+\frac{i (r+s+1)}{2}) 
 \mathcal{F}(y+\frac{i (r+s+1)}{2})}
 {\tan \eta (v-y-\frac{i(r+s+2)}{2})
 \mathcal{T}^{(r+s+1)}_{1}(y+\frac{i(r+s)}{2})}
 \nonumber \\
&& \hspace{20pt}+
\oint_{\overline{C}^{(r+s+1)}_{1}} \frac{{\mathrm d} y}{2\pi i} 
 \frac{\eta 
 \mathcal{T}^{(r+s)}_{1}(y-\frac{i (r+s+1)}{2}) 
 \mathcal{F}(y-\frac{i (r+s+1)}{2})}
 {\tan \eta (v-y+\frac{i(r+s+2)}{2})
 \mathcal{T}^{(r+s+1)}_{1}(y-\frac{i(r+s)}{2})} ,
\label{nlie-generalb}
 \\
&& \hspace{20pt} 
\mathcal{F}(v)=\lim_{N \to \infty }\widetilde{T}^{(r+s+2)}_{1}(v)=
\frac{
A_{1}(v)-
\zeta 
A_{2}(v)
}
{(-1)^{s}A_{3}(v)+(-1)^{s} 
\zeta 
A_{4}(v)},
\label{det-hashi}
\end{eqnarray}
where 
\begin{eqnarray}
&& A_{1}(v)=\det _{1\le j,k \le s+2}
\left(f_{j,k}
\left(
v-\frac{j+k-s-3}{2}i
\right) 
\right) \\
&& \quad f_{j,k}(v)=\mathcal{T}^{(r+1+j-k)}_{1}(v) 
 \quad \mbox{for} \quad (j,k) \ne (s+2,1), 
\quad  f_{s+2,1}(v)=0,  \nonumber \\
&& A_{2}(v)=
\det _{1\le j,k \le s+1}
\left(g_{j,k}
\left(
v-\frac{j+k-s-2}{2}i
\right) 
\right) 
 \\
&& \quad g_{j,k}(v)=\mathcal{T}^{(r+2+j-k)}_{1}(v) 
 \quad \mbox{for} \quad (j,k) \ne (s+1,1), 
\quad  g_{s+1,1}(v)=0,  \nonumber \\ 
&& A_{3}(v)=\det _{1\le j,k \le s+1}
\left(\mathcal{T}^{(r+j-k)}_{1}
\left(
v-\frac{j+k-2-s}{2}i
\right) 
\right), \\
&& A_{4}(v)=
\det _{1\le j,k \le s}
\left(\mathcal{T}^{(r+j-k+1)}_{1}
\left(
v-\frac{j+k-s-1}{2}i
\right) 
\right) 
.
\end{eqnarray}
If $s=-1$, then $A_{1}(v)=A_{4}(v)=0$ and 
$A_{2}(v)=A_{3}(v)=1$, and consequently   
(\ref{det-hashi}) reduces to 
${\mathcal F}(v)=\mathcal{T}^{(r+1)}_{1}(v)=Q^{(r+1)}_{1}=
\zeta =
e^{\frac{\mu_{1}+\cdots +\mu_{r+1}}{T}}$, where 
the determinants should be interpreted as 
$\det_{1\le j,k \le 0} (\cdots )=1$, $\det_{1\le j,k \le -1} (\cdots )=0$.  Thus   
(\ref{nlie-general}) and (\ref{nlie-generalb}) 
reduce to the NLIE for $U_{q}(\widehat{sl}(r+1))$ in \cite{TT05}.
In particular for $s=0$ ($U_{q}(\widehat{sl}(r+1|1))$ case, we can use 
(\ref{sol}):  
\begin{eqnarray}
&& {\mathcal T}^{(a)}_{1}(v)=Q^{(a)}_{1} 
+
\oint_{C^{(a)}_{1}} \frac{{\mathrm d} y}{2\pi i} 
 \frac{\eta 
 \mathcal{T}^{(a-1)}_{1}(y+\frac{i a}{2}) 
 \mathcal{T}^{(a+1)}_{1}(y+\frac{i a}{2})}
 {\tan \eta (v-y-\frac{i(a+1)}{2})
 \mathcal{T}^{(a)}_{1}(y+\frac{i(a-1)}{2})}
 \nonumber \\
&& \hspace{76pt} +
\oint_{\overline{C}^{(a)}_{1}} \frac{{\mathrm d} y}{2\pi i} 
 \frac{\eta 
 \mathcal{T}^{(a-1)}_{1}(y-\frac{i a}{2}) 
 \mathcal{T}^{(a+1)}_{1}(y-\frac{i a}{2})}
 {\tan \eta (v-y+\frac{i(a+1)}{2})
 \mathcal{T}^{(a)}_{1}(y-\frac{i(a-1)}{2})}
 \nonumber \\ 
 && \hspace{100pt} 
 {\rm for} \quad a \in \{1,2,\dots r \},
 \label{nlie-s=0} \\
&& {\mathcal T}^{(r+1)}_{1}(v)=Q^{(r+1)}_{1} 
 \nonumber \\ 
&& \hspace{10pt}+
\oint_{C^{(r+1)}_{1}} \frac{{\mathrm d} y}{2\pi i} 
 \frac{\eta 
 \mathcal{T}^{(r)}_{1}(y+\frac{i (r+1)}{2}) 
 \mathcal{T}^{(r+1)}_{1}(y+\frac{i(r+2)}{2})}
 {\tan \eta (v-y-\frac{i(r+2)}{2})
 (
 \zeta 
+\mathcal{T}^{(r)}_{1}(y+\frac{i(r+1)}{2}))}
 \nonumber \\
&& \hspace{10pt}+
\oint_{\overline{C}^{(r+1)}_{1}} \frac{{\mathrm d} y}{2\pi i} 
 \frac{\eta 
 \mathcal{T}^{(r)}_{1}(y-\frac{i (r+1)}{2}) 
 \mathcal{T}^{(r+1)}_{1}(y-\frac{i(r+2)}{2})}
 {\tan \eta (v-y+\frac{i(r+2)}{2})
 (
 \zeta 
+\mathcal{T}^{(r)}_{1}(y-\frac{i(r+1)}{2}))}.
 \nonumber \\
 && \label{nlie-s=0b}
\end{eqnarray}
The free energy per site is given by a solution of these 
NLIE (\ref{nlie-general})-(\ref{nlie-s=0b})
\begin{eqnarray}
f=J \cosh \eta -T \log \mathcal{T}^{(1)}_{1}(0).
 \label{free-en}
\end{eqnarray}
In these NLIE (\ref{nlie-general})-(\ref{nlie-s=0b}), 
the number of unknown functions and equations is 
$r+s+1$, which contrasts with TBA equations \cite{Sch87,Sch92,EK94,JKS98,Sa99}.
\subsection{The nonlinear integral equations for $\xi=-1$}
Next,  
let us consider the NLIE (\ref{nlie2}) for $\xi=-1$ and $a=1$. 
Taking note on the fact ${\widetilde T}^{(0)}_{m}(v)=1$ (cf.(\ref{0m})), 
we can drop the second terms in the two brackets $\{\cdots \}$ in (\ref{nlie2}) 
since they have no poles at $y=0$.
Then the NLIE (\ref{nlie2}) reduce to the following NLIE on 
${\mathcal T}^{(1)}_{m}(v)=\lim_{N \to \infty}\widetilde{T}^{(1)}_{m}(v)$ 
 after the Trotter limit $N \to \infty $ with $u=-\frac{J \sinh \eta }{\eta N T}$.
\begin{eqnarray}
{\mathcal T}^{(1)}_{m}(v)=Q^{(1)}_{m} 
&+&
\oint_{C^{(1)}_{m}} \frac{{\mathrm d} y}{2\pi i} 
 \frac{\eta 
 \mathcal{T}^{(1)}_{m-1}(y-\frac{i m}{2}) 
 \mathcal{T}^{(1)}_{m+1}(y-\frac{i m}{2})}
 {\tan \eta (v-y+\frac{i(m+1)}{2})
 \mathcal{T}^{(1)}_{m}(y-\frac{i(m-1)}{2})}
 \nonumber \\
&+&
\oint_{\overline{C}^{(1)}_{m}} \frac{{\mathrm d} y}{2\pi i} 
 \frac{\eta 
 \mathcal{T}^{(1)}_{m-1}(y+\frac{i m}{2}) 
 \mathcal{T}^{(1)}_{m+1}(y+\frac{i m}{2})}
 {\tan \eta (v-y-\frac{i(m+1)}{2})
 \mathcal{T}^{(1)}_{m}(y+\frac{i(m-1)}{2})}
 \nonumber \\ 
 && \hspace{70pt} 
 {\rm for} \quad m \in {\mathbb Z}_{\ge 1},
 \label{infinitenlie-xi=-1}
\end{eqnarray}
where 
\begin{eqnarray} 
\mathcal{T}^{(1)}_{0}(v)=
\lim_{N \to \infty} \widetilde{T}^{(1)}_{0}(v)
=\exp \left(-\frac{2J (\sinh \eta)^{2} }
{T(\cosh \eta -\cos (2\eta v))}\right),
\end{eqnarray}
and the contour $C^{(1)}_{m}$ (resp. $\overline{C}^{(1)}_{m}$) 
is a counterclockwise closed loop around $y=0$ (resp. $y=0$) 
which satisfies the condition 
$y \ne v+\frac{m+1}{2}i+\frac{\pi n}{\eta}$ 
(resp. $y \ne v-\frac{m+1}{2}i+\frac{\pi n}{\eta}$) and 
does not surround 
$z^{(1)}_{m}+\frac{m-1}{2}i+\frac{\pi n}{\eta}$, 
$(1+m)i
+\frac{\pi n}{\eta}$, $\frac{\pi k}{\eta}$ 
(resp. 
$z^{(1)}_{m}-\frac{m-1}{2}i+\frac{\pi n}{\eta}$, 
$-(1+m)i +\frac{\pi n}{\eta}$, $\frac{\pi k}{\eta}$) 
 ($n \in \mathbb{Z}$, $k \in \mathbb{Z}-\{0\}$). 
Here $\{z^{(1)}_{m}\}$ are zeros of ${\mathcal T}^{(1)}_{m}(v)$: 
${\mathcal T}^{(1)}_{m}(z^{(1)}_{m})=0$. 
These are an infinite number of coupled NLIE. 
We can reduce them as $\xi=1$ case. 
By using (\ref{a+r+s-b}) in the limit $N \to \infty$,
 we can reduce (\ref{infinitenlie-xi=-1}) 
as follows, 
where $Z^{(a)}_{m}(v)=1$ for $\xi=-1$.
\begin{eqnarray}
{\mathcal T}^{(1)}_{m}(v)=Q^{(1)}_{m} 
&+&
\oint_{C^{(1)}_{m}} \frac{{\mathrm d} y}{2\pi i} 
 \frac{\eta 
 \mathcal{T}^{(1)}_{m-1}(y-\frac{i m}{2}) 
 \mathcal{T}^{(1)}_{m+1}(y-\frac{i m}{2})}
 {\tan \eta (v-y+\frac{i(m+1)}{2})
 \mathcal{T}^{(1)}_{m}(y-\frac{i(m-1)}{2})}
 \nonumber \\
&+&
\oint_{\overline{C}^{(1)}_{m}} \frac{{\mathrm d} y}{2\pi i} 
 \frac{\eta 
 \mathcal{T}^{(1)}_{m-1}(y+\frac{i m}{2}) 
 \mathcal{T}^{(1)}_{m+1}(y+\frac{i m}{2})}
 {\tan \eta (v-y-\frac{i(m+1)}{2})
 \mathcal{T}^{(1)}_{m}(y+\frac{i(m-1)}{2})}
 \nonumber \\ 
 && \hspace{70pt} 
 {\rm for} \quad m \in \{1,2,\dots r+s \},
 \label{nlie-xi=-1} \\
{\mathcal T}^{(1)}_{r+s+1}(v)=Q^{(1)}_{r+s+1} 
&+&
\oint_{C^{(1)}_{r+s+1}} \frac{{\mathrm d} y}{2\pi i} 
 \frac{\eta 
 \mathcal{T}^{(1)}_{r+s}(y-\frac{i (r+s+1)}{2}) 
 \mathcal{G}(y-\frac{i (r+s+1)}{2})}
 {\tan \eta (v-y+\frac{i(r+s+2)}{2})
 \mathcal{T}^{(1)}_{r+s+1}(y-\frac{i(r+s)}{2})}
 \nonumber \\
&& \hspace{-70pt}+
\oint_{\overline{C}^{(1)}_{r+s+1}} \frac{{\mathrm d} y}{2\pi i} 
 \frac{\eta 
 \mathcal{T}^{(1)}_{r+s}(y+\frac{i (r+s+1)}{2}) 
 \mathcal{G}(y+\frac{i (r+s+1)}{2})}
 {\tan \eta (v-y-\frac{i(r+s+2)}{2})
 \mathcal{T}^{(1)}_{r+s+1}(y+\frac{i(r+s)}{2})} ,
 \label{nlie-xi=-1b}
\end{eqnarray}
%
\begin{eqnarray}
\mathcal{G}(v)=
\lim_{N \to \infty}
\widetilde{T}^{(1)}_{r+s+2}(v)=
\frac{
\zeta 
A_{5}(v)-A_{6}(v)
}
{(-1)^{r}
\zeta 
A_{7}(v)+(-1)^{r} A_{8}(v)},
\end{eqnarray}
where 
\begin{eqnarray}
&& A_{5}(v)=\det _{1\le j,k \le r+2}
\left( h_{j,k}
\left(
v-\frac{r+3-j-k}{2}i
\right) 
\right) \\
&& \quad h_{j,k}(v)={\mathcal T}^{(1)}_{s+1+j-k}(v) 
 \quad \mbox{for} \quad (j,k) \ne (2+r,1), 
\quad  h_{r+2,1}(v)=0,  \nonumber \\
&& A_{6}(v)=
\det _{1\le j,k \le r+1}
\left( b_{j,k}
\left(
v-\frac{r+2-j-k}{2}i
\right) 
\right) 
 \\
&& \quad b_{j,k}(v)={\mathcal T}^{(1)}_{s+2+j-k}(v) 
 \quad \mbox{for} \quad (j,k) \ne (r+1,1), 
\quad  b_{r+1,1}(v)=0,  \nonumber \\ 
&& A_{7}(v)=\det _{1\le j,k \le r+1}
\left({\mathcal T}^{(1)}_{s+j-k}
\left(
v-\frac{r+2-j-k}{2}i
\right) 
\right), \\
&&A_{8}(v)=
\det _{1\le j,k \le r}
\left({\mathcal T}^{(1)}_{s+1+j-k}
\left(
v-\frac{r+1-j-k}{2}i
\right) 
\right) 
,
\end{eqnarray}
where ${\mathcal T}^{(1)}_{m}(v)=0$ for $m<0 $.
%

In particular for $r=0$ ($U_{q}(\widehat{sl}(1|s+1))$ case, we can use 
(\ref{sol}):  
\begin{eqnarray}
&& {\mathcal T}^{(1)}_{m}(v)=Q^{(1)}_{m} +
\oint_{C^{(1)}_{m}} \frac{{\mathrm d} y}{2\pi i} 
 \frac{\eta 
 \mathcal{T}^{(1)}_{m-1}(y-\frac{i m}{2}) 
 \mathcal{T}^{(1)}_{m+1}(y-\frac{i m}{2})}
 {\tan \eta (v-y+\frac{i(m+1)}{2})
 \mathcal{T}^{(1)}_{m}(y-\frac{i(m-1)}{2})}
 \nonumber \\
&& \hspace{76pt} +
\oint_{\overline{C}^{(1)}_{1}} \frac{{\mathrm d} y}{2\pi i} 
 \frac{\eta 
 \mathcal{T}^{(1)}_{m-1}(y+\frac{i m}{2}) 
 \mathcal{T}^{(1)}_{m+1}(y+\frac{i m}{2})}
 {\tan \eta (v-y-\frac{i(m+1)}{2})
 \mathcal{T}^{(1)}_{m}(y+\frac{i(m-1)}{2})}
 \nonumber \\ 
 && \hspace{100pt} 
 {\rm for} \quad m \in \{1,2,\dots s \},
 \label{nlie-r=0} \\
&& {\mathcal T}^{(1)}_{s+1}(v)=Q^{(1)}_{s+1} 
\nonumber \\
&& \hspace{8pt} +
\oint_{C^{(1)}_{s+1}} \frac{{\mathrm d} y}{2\pi i} 
 \frac{\eta 
 \mathcal{T}^{(1)}_{s}(y-\frac{i (s+1)}{2}) 
 \mathcal{T}^{(1)}_{s+1}(y-\frac{i(s+2)}{2})}
 {\tan \eta (v-y+\frac{i(s+2)}{2})
 (
 \zeta^{-1}
+\mathcal{T}^{(1)}_{s}(y-\frac{i(s+1)}{2}))}
 \nonumber \\
&& \hspace{8pt}+
\oint_{\overline{C}^{(1)}_{s+1}} \frac{{\mathrm d} y}{2\pi i} 
 \frac{\eta 
 \mathcal{T}^{(1)}_{s}(y+\frac{i (s+1)}{2}) 
 \mathcal{T}^{(1)}_{s+1}(y+\frac{i(s+2)}{2})}
 {\tan \eta (v-y-\frac{i(s+2)}{2})
 (
 \zeta^{-1}
+\mathcal{T}^{(1)}_{s}(y+\frac{i(s+1)}{2}))}. 
\nonumber \\ 
\label{nlie-r=0b}
\end{eqnarray}
The free energy per site is given by a solution of these 
NLIE (\ref{nlie-xi=-1})-(\ref{nlie-r=0b})
\begin{eqnarray}
f=-J \cosh \eta -T \log \mathcal{T}^{(1)}_{1}(0).
 \label{free-en2}
\end{eqnarray}
In some sense, these NLIE are \symbol{"60}dual' to the ones in the previous section. 
The NLIE (\ref{nlie-xi=-1})-(\ref{nlie-r=0b})have only $r+s+1$ unknown functions. 
These NLIE have never been considered before even for $U_{q}(\widehat{sl}(2))$ case. 
\section{High temperature expansions} 
In this section, we will calculate the high temperature 
expansion of the free energy from our new NLIE. 
For large $T/|J|$, we assume the following expansion :
\begin{eqnarray}
&&\mathcal{T}^{(a)}_{1}(v)=
 \exp \left(\sum_{n=0}^{{\mathrm deg}}b_{n}^{(a)}(v)(\frac{J}{T})^{n} 
+O((\frac{J}{T})^{{\mathrm deg}+1}) \right)
 \nonumber 
\\
&& =Q^{(a)}_{1}\Biggl\{ 1+b^{(a)}_{1}(v)\frac{J}{T}+
\left(b^{(a)}_{2}(v)+\frac{(b^{(a)}_{1}(v))^2}{2}\right)(\frac{J}{T})^2
+ \label{hte-ta} \\
&& \left(b^{(a)}_{3}(v)+b^{(a)}_{2}(v)b^{(a)}_{1}(v)+
\frac{(b^{(a)}_{1}(v))^3}{6}\right)
(\frac{J}{T})^3 +\cdots \Biggr\}+O((\frac{J}{T})^{{\mathrm deg}+1}),
\nonumber 
\end{eqnarray}
where $b_{0}^{(a)}(v)=\log Q^{(a)}_{1}$. 
Here we do not expand $\{Q^{(b)}_{1}\}_{b \ge 1}$ with respect to $\frac{J}{T}$. 
Thus the coefficients $\{b^{(a)}_{n}(v) \}$ 
themselves depend on $\frac{1}{T}$.
In this sense, our high temperature expansion formula 
 is different from ordinary one. 
Substituting this (\ref{hte-ta}) into some of the NLIE 
(\ref{nlie4})-(\ref{nlie-s=0b}), 
we can calculate the coefficients $\{b^{(a)}_{n}(v) \}$ up to the order of $n={\mathrm deg}$. 
Note that we only need $\{b^{(1)}_{n}(0) \}$ to calculate the free energy (\ref{free-en}).  
Taking note on this fact, 
firstly we use
\footnote{As for numerical calculations of the free energy, 
we expect that the reduced NLIE (\ref{nlie-general})-(\ref{nlie-s=0b}) 
are easier to use than the non-reduced NLIE (\ref{nlie4}).}
 a subset (NLIE for $a \in \{1,2,\dots, {\mathrm deg} \}$) 
of the non-reduced NLIE (\ref{nlie4}) 
rather than the reduced NLIE (\ref{nlie-general})-(\ref{nlie-s=0b}). 
We have observed that $b^{(1)}_{n}(0)$ can be expressed in terms of 
\footnote{For $s=-1$ case, 
they are 
$Q^{(1)}_{1},Q^{(2)}_{1}, \dots ,Q^{(d)}_{1}$: 
$d=\min (n+1,r+1)$ since 
$Q^{(a)}_{1}=0$ if $a \ge r+2$.}
$Q^{(1)}_{1},Q^{(2)}_{1}, \dots ,Q^{(n+1)}_{1}$.  
We have calculated the coefficients by using Mathematica. 
As examples, we shall enumerate the coefficients $\{b^{(1)}_{n}(0) \}$ up to the 
order of $5$, where we put $\Delta=\cosh \eta $. 
\begin{eqnarray}
&& \hspace{-20pt}
b^{(1)}_{1}(0)= \frac{2 \Delta Q^{(2)}_{1}}{{Q^{(1)}_{1}}^2}, 
 \label{coe1} \\
&& \hspace{-20pt}
b^{(1)}_{2}(0)=-\frac{6 \Delta^2 {Q^{(2)}_{1}}^2}{{Q^{(1)}_{1}}^4}+\frac{\left(2 \Delta^2+1\right)
   Q^{(2)}_{1}}{{Q^{(1)}_{1}}^2}+\frac{\left(4 \Delta^2-1\right) Q^{(3)}_{1}}{{Q^{(1)}_{1}}^3},
 \label{coe2} \\ 
&& \hspace{-20pt}
b^{(1)}_{3}(0)=\frac{80 {Q^{(2)}_{1}}^3 \Delta^3}{3
   {Q^{(1)}_{1}}^6}
+\frac{8 Q^{(3)}_{1} \Delta^3}{{Q^{(1)}_{1}}^3}
+\frac{\left(\frac{4 \Delta^3}{3}+2 \Delta\right)
   Q^{(2)}_{1}}{{Q^{(1)}_{1}}^2}
\nonumber \\
&& 
\hspace{-15pt}
+\frac{\left(8 \Delta-32 \Delta^3\right) Q^{(2)}_{1} Q^{(3)}_{1}}{{Q^{(1)}_{1}}^5}
+\frac{\left(-12 \Delta^3-6
   \Delta\right) {Q^{(2)}_{1}}^2
+\left(8 \Delta^3-4 \Delta\right) Q^{(4)}_{1}}{{Q^{(1)}_{1}}^4},
 \label{coe3} \\
&&\hspace{-20pt}
 b^{(1)}_{4}(0)=-\frac{140 \Delta^4
   {Q^{(2)}_{1}}^4}{{Q^{(1)}_{1}}^8}
+\frac{\left(240 \Delta^4-60 \Delta^2\right) Q^{(3)}_{1}
   {Q^{(2)}_{1}}^2}{{Q^{(1)}_{1}}^7}
\nonumber \\
&& 
+\frac{\left(\frac{2 \Delta^4}{3}+2 \Delta^2+\frac{1}{4}\right)
   Q^{(2)}_{1}}{{Q^{(1)}_{1}}^2}
+\frac{\left(\frac{28 \Delta^4}{3}+\frac{14 \Delta^2}{3}-\frac{1}{4}\right)
   Q^{(3)}_{1}}{{Q^{(1)}_{1}}^3}
\nonumber \\
&& 
+\frac{\left(-14 \Delta^4-\frac{56 \Delta^2}{3}-\frac{3}{2}\right) 
 {Q^{(2)}_{1}}^2+\left(24 \Delta^4-8
   \Delta^2-1\right) Q^{(4)}_{1}}{{Q^{(1)}_{1}}^4}
\nonumber \\
&& 
+\frac{\left(80 \Delta^4+40 \Delta^2\right) {Q^{(2)}_{1}}^3+\left(40 \Delta^2-80 \Delta^4\right)
   Q^{(4)}_{1} Q^{(2)}_{1}}{{Q^{(1)}_{1}}^6}
\nonumber \\
&& 
+\frac{\left(-40 \Delta^4+20 \Delta^2-\frac{5}{2}\right) {Q^{(3)}_{1}}^2}{{Q^{(1)}_{1}}^6}
\nonumber \\
&& 
+\frac{\left(-96 \Delta^4-8
   \Delta^2+4\right) Q^{(2)}_{1} Q^{(3)}_{1}
 +\left(16 \Delta^4-12 \Delta^2+1\right) Q^{(5)}_{1}}{{Q^{(1)}_{1}}^5},
\label{coe4} 
\end{eqnarray}
\begin{eqnarray}
&& \hspace{-15pt} b^{(1)}_{5}(0)=\frac{4032 \Delta^5
   {Q^{(2)}_{1}}^5}{5 {Q^{(1)}_{1}}^{10}}
 +\frac{\left(448 \Delta^3-1792 \Delta^5\right) Q^{(3)}_{1}
   {Q^{(2)}_{1}}^3}{{Q^{(1)}_{1}}^9}
\nonumber \\
&& 
 +\frac{\left(\frac{4 \Delta^5}{15}+\frac{4 \Delta^3}{3}+\frac{\Delta}{2}\right)
   Q^{(2)}_{1}}{{Q^{(1)}_{1}}^2}
+\frac{\left(8 \Delta^5+10 \Delta^3+\frac{\Delta}{2}\right) Q^{(3)}_{1}}{{Q^{(1)}_{1}}^3}
\nonumber \\
&& 
 +\frac{\left(-12
   \Delta^5-30 \Delta^3-8 \Delta\right) {Q^{(2)}_{1}}^2+\left(40 \Delta^5-6 \Delta\right)
  Q^{(4)}_{1}}{{Q^{(1)}_{1}}^4}
\nonumber \\
&& 
+\frac{\left(-560 \Delta^5-280
   \Delta^3\right) {Q^{(2)}_{1}}^4+\left(672 \Delta^5-336 \Delta^3\right) 
 Q^{(4)}_{1} {Q^{(2)}_{1}}^2}{{Q^{(1)}_{1}}^8}
 \nonumber \\
&& 
+\frac{\left(672 \Delta^5-336 \Delta^3+42 \Delta\right)
   {Q^{(3)}_{1}}^2 Q^{(2)}_{1}}{{Q^{(1)}_{1}}^8}
\nonumber \\
&& 
+\frac{\left(-160 \Delta^5-100 \Delta^3+11 \Delta\right) Q^{(2)}_{1} Q^{(3)}_{1}
 +\left(64 \Delta^5-40
   \Delta^3\right) Q^{(5)}_{1}}{{Q^{(1)}_{1}}^5}
\nonumber \\
&& 
\hspace{-10pt}
+\frac{\left(960 \Delta^5+120 \Delta^3-60 \Delta\right) Q^{(3)}_{1} {Q^{(2)}_{1}}^2+\left(-192
   \Delta^5+144 \Delta^3-12 \Delta\right) Q^{(5)}_{1} Q^{(2)}_{1}}{{Q^{(1)}_{1}}^7}
\nonumber \\
&&
+\frac{\left(-192 \Delta^5+144 \Delta^3-24 \Delta\right) Q^{(3)}_{1}
   Q^{(4)}_{1}}{{Q^{(1)}_{1}}^7}
\nonumber \\
&& 
+\frac{\left(\frac{400 \Delta^5}{3}+\frac{500 \Delta^3}{3}+20 \Delta\right) {Q^{(2)}_{1}}^3+\left(-320
   \Delta^5+80 \Delta^3+30 \Delta\right) Q^{(4)}_{1} Q^{(2)}_{1}}{{Q^{(1)}_{1}}^6}
\nonumber \\
&& 
+\frac{\left(40 \Delta^3-160 \Delta^5\right) {Q^{(3)}_{1}}^2+\left(32 \Delta^5-32 \Delta^3+6
   \Delta\right) Q^{(6)}_{1}}{{Q^{(1)}_{1}}^6}.
   \label{coe5} 
\end{eqnarray}
In deriving these coefficients (\ref{coe1})-(\ref{coe5}), we 
did not assume (\ref{limit}). Of course, when one calculate the free energy of the model, 
 one must assume (\ref{limit}) and (\ref{para}). 
We can also rewrite the coefficient $b^{(1)}_{n}(0)$ in terms of 
$Q^{(1)}_{1},Q^{(2)}_{1},\dots,Q^{(d)}_{1}$ and $\zeta$ 
\footnote{
$Q^{(r+1)}_{1}=\zeta$ if $s=-1$.}
( $d=\min (n+1,r+s+1)$ ) since $Q^{(a)}_{1}$ for $a \in {\mathbb Z}_{\ge r+s+2}$ can 
be written in terms of $Q^{(1)}_{1},Q^{(2)}_{1},\dots,Q^{(r+s+1)}_{1}$ and $\zeta$ 
due to the relation (\ref{a+r+s}) in the limit $v \to i\eta^{-1} \infty $ 
(see also an example: (\ref{Q11-sl21})-(\ref{Qa1-sl21})). 
If $b^{(n)}_{1}(0)$ is written in terms of 
$Q^{(1)}_{1},Q^{(2)}_{1},\dots,Q^{(d)}_{1}$ and $\zeta$ ( 
$d=\min (n+1,r+s+1)$), it should be the coefficient 
of the high temperature expansion directly derived from 
 the reduced NLIE (\ref{nlie-general})-(\ref{nlie-s=0b}). 
 Of course these two expressions of the coefficient $b^{(1)}_{n}(0)$ are 
 equivalent under the relations (\ref{limit}) and (\ref{para}). 
 
\begin{figure}
\begin{center}
\includegraphics[width=1\textwidth]
{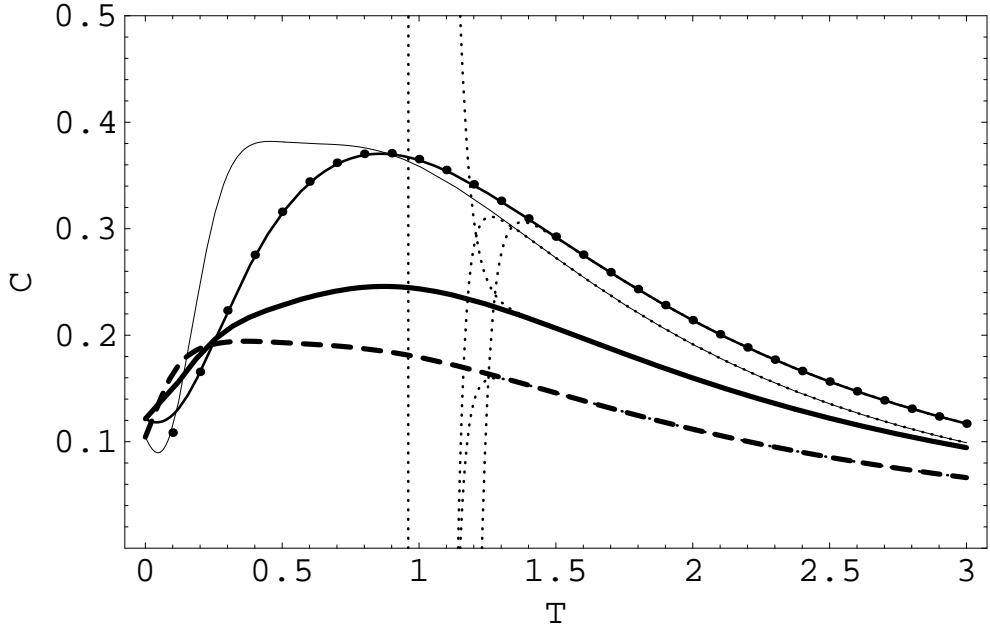}
\end{center}
\caption{Temperature dependence of the high temperature 
expansion of the specific heat $C$ 
for the rank 2 case ($r+s=1$, $J=1$, $q=1$,  
$\mu_{a}=0$ ($a \in B$)). We have plotted 
plan series (dotted lines) of $C$ in Appendix and their Pade approximations 
of order [$n$,$d$] (numerator: a degree $n$ polynomial of $1/T$, 
 denominator: a degree $d$ polynomial of $1/T$) 
by using Mathematica:  
 each line denotes $C$ for 
$sl(3|0)$ with [20,20] (thin), $sl(2|1)$ with [17,17] (medium),
$sl(1|2)$ with [17,17] (thick), $sl(0|3)$ [20,20] (dashed thick) respectively. 
We have also plotted (thick dots) 
a result of numerical calculation from another NLIE by J\"uttner 
and Kl\"umper \cite{JK97} for the $sl(2|1)$ case. 
 C for the $sl(3|0)$ case was also 
considered in \cite{FK02,FK99}.}
\label{specific2}
\end{figure}

\begin{figure}
\begin{center}
\includegraphics[width=1\textwidth]
{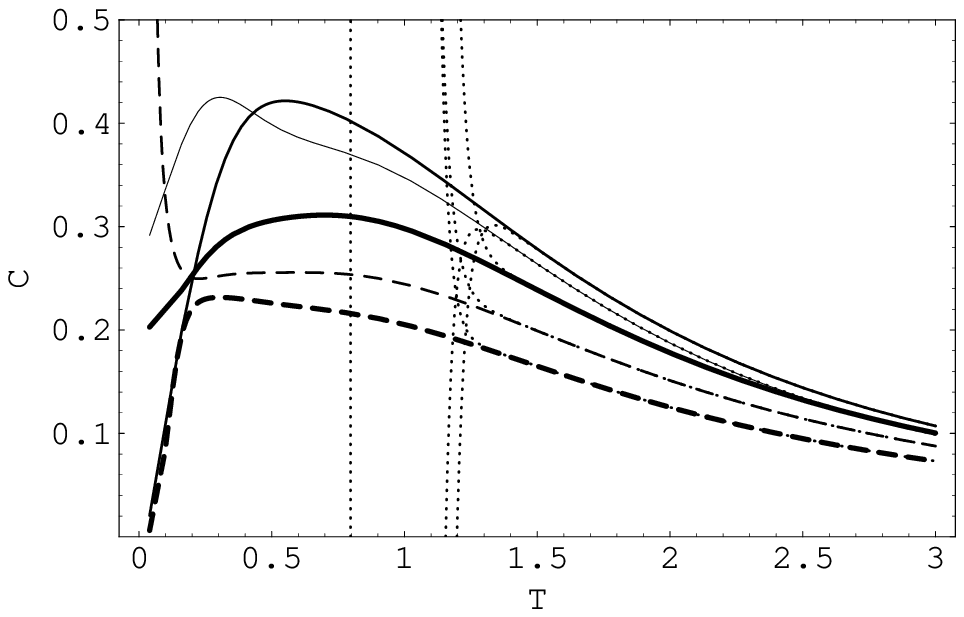}
\end{center}
\caption{Temperature dependence of the high temperature 
expansion of the specific heat $C$ 
for the rank 3 case ($r+s=2$, $J=1$, $q=1$,  
$\mu_{a}=0$ ($a \in B$)). We have plotted 
plan series (dotted lines) of $C$ in Appendix and their Pade approximations 
of order [$n$,$d$] (numerator: a degree $n$ polynomial of $1/T$, 
 denominator: a degree $d$ polynomial of $1/T$):  
 each line denotes $C$ for 
$sl(4|0)$ with [19,20] (thin), $sl(3|1)$ with [17,17] (medium),
$sl(2|2)$ with [16,16] (thick), $sl(1|3)$ with [17,17] 
(dashed medium), $sl(0|4)$ with [18,21] (dashed thick) respectively. 
 C for the $sl(4|0)$ case was also 
considered in \cite{FK02}.}
\label{specific3}
\end{figure}

\begin{figure}
\begin{center}
\includegraphics[width=1\textwidth]
{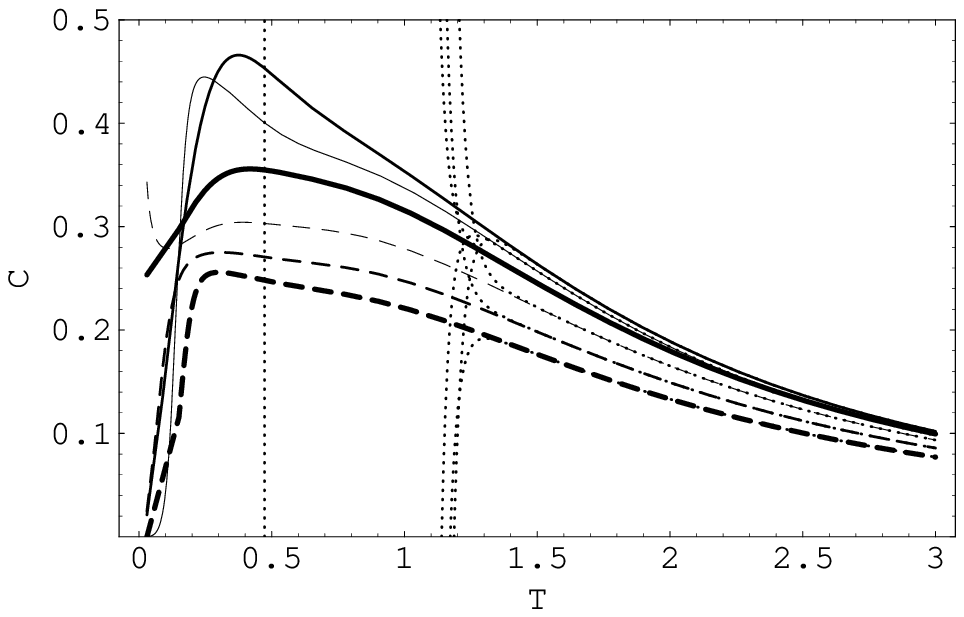}
\end{center}
\caption{Temperature dependence of the high temperature 
expansion of the specific heat $C$ 
for the rank 4 case ($r+s=3$, $J=1$, $q=1$,  
$\mu_{a}=0$ ($a \in B$)). We have plotted 
plan series (dotted lines) of $C$ in Appendix and their Pade approximations 
of order [$n$,$d$] (numerator: a degree $n$ polynomial of $1/T$, 
 denominator: a degree $d$ polynomial of $1/T$):  
 each line denotes $C$ for 
$sl(5|0)$ with [17,21] (thin), $sl(4|1)$ with [16,18] (medium),
$sl(3|2)$ with [17,17] (thick), 
$sl(2|3)$ with [16,17] (dashed thin), $sl(1|4)$ with [16,18] 
(dashed medium), $sl(0|5)$ with [17,21] (dashed thick) respectively. }
\label{specific4}
\end{figure}

 For fixed values of parameters, we have calculated 
 the high temperature expansion for much higher order (see, appendix). 
We have plotted the high temperature expansion 
of the specific heat (Figure \ref{specific2}-\ref{specific4}).
 Here we have adopted the Pade approximation method. 
There is a duality among the specific heats with respect to interchange of 
$r$ and $s$. 
In particular, $r=s$ case is self-dual, then 
the specific heat becomes an even function of $T$ (see (\ref{hte-sl22})). 
In Figure \ref{specific2}, we have also plotted a result of 
 a numerical calculation by another NLIE \cite{JK97}.
We find a good agreement 
 between our result and their result except for very low temperature region. 
 
We can also calculate the high temperature expansion from the NLIE 
for $\xi=-1$ in subsection 3.2. 
Similar to $\xi=1$ case, we assume 
\begin{eqnarray}
&&\mathcal{T}^{(1)}_{m}(v)=
 \exp \left(\sum_{n=0}^{{\mathrm deg}}\widehat{b}_{m,n}(v)(\frac{J}{T})^{n} 
+O((\frac{J}{T})^{{\mathrm deg}+1}) \right) ,
\label{hte-tm}
\end{eqnarray}
where $\widehat{b}_{m,0}(v)=\log Q^{(1)}_{m}$. 
Here we do not expand $\{ Q^{(1)}_{k} \}_{k \ge 1}$ with respect to  $\frac{J}{T}$.  
(\ref{hte-ta}) for $a=1$ should coincide with 
(\ref{hte-tm}) for $m=1$  up to a factor from 
the normalization function (\ref{normal}). 
Thus we have 
\begin{eqnarray}
b^{(1)}_{n}(0)=\widehat{b}_{1,n}(0)+2\Delta \delta_{n,1}
 \label{ty1}
\end{eqnarray}
Due to symmetry between the NLIE for $\xi=1$ and the one for $\xi=-1$, 
the following relation follows:
\begin{eqnarray}
\widehat{b}_{1,n}(0)=(-1)^{n}b^{(1)}_{n}(0)|_{Q^{(a)}_{1} \to Q^{(1)}_{a} 
 \ {\rm for} \ a \ge 1}.
 \label{ty2}
\end{eqnarray}
For example, (\ref{ty1}) and  (\ref{ty2}) for $n=1$ 
and (\ref{coe1}) reproduce 
 the $Q$-system (\ref{Q-sys}) for $(a,m)=(1,1)$. 
From the relations 
(\ref{ty1}) and  (\ref{ty2}) for $n=2$ and (\ref{coe2}), we obtain 
 identities among characters 
\begin{eqnarray} 
&& \hspace{-40pt} 
-3 {Q^{(2)}_{1}}^{2}+Q^{(2)}_{1}{Q^{(1)}_{1}}^{2}+2 Q^{(3)}_{1}Q^{(1)}_{1}
=-3 {Q^{(1)}_{2}}^{2}+Q^{(1)}_{2}{Q^{(1)}_{1}}^{2}+2 Q^{(1)}_{3}Q^{(1)}_{1}, \\
&& \hspace{-40pt}
 Q^{(2)}_{1}Q^{(1)}_{1}-Q^{(3)}_{1}=Q^{(1)}_{2}Q^{(1)}_{1}-Q^{(1)}_{3},
\end{eqnarray}
where we have used the fact that $Q^{(a)}_{m}$ does not depend on $\Delta $. 
These relations can be proved  from the  
relations (\ref{jacobi-trudi}), (\ref{jacobi-trudi2}) and (\ref{limit}).

Some comments on references on the high temperature expansion 
are in order. 
The high temperature expansion of the free energy was
 calculated from the Takahashi's NLIE for 
 the $XXX$-model up to the order of 100 \cite{ShT02}; 
 the $XXZ$-model up to the order of 99 \cite{TT05}. 
As for the higher rank or higher spin case, we have some results
 \cite{T02,T03,T04,TT05} from NLIE. 
In particular, our result on the $sl(r+1)$ Uimin-Sutherland model 
in \cite{T03} was applied \cite{BGOSTF03,YRFC04,YRZ04,BGO04,BGOF04,BGOT05} 
to spin ladder models and 
good agreement
was seen between theoretical results and 
experimental data. 
We note that 
the coefficients (\ref{coe1})-(\ref{coe3}) coincide with eqs. 
(4.14)-(4.16) in \cite{TT05}. 
Note however that the coefficients in our paper are more general than the ones in 
\cite{TT05} since the value of $Q^{(a)}_{1}$ (\ref{limit})
 was restricted to $s=-1$ case in \cite{TT05}. 
There are also several works on high temperature expansions by different methods 
(see for example, \cite{DV95,RST02,BEU00,FK02,F03}).
\section{Concluding remarks}
In this paper, we have derived NLIE which contain only $r+s+1$ unknown functions 
for the $U(\widehat{sl}(r+1|s+1))$ Perk-Schultz model.  
The key is a duality for the auxiliary function (\ref{dual}) 
and the quantum (supersymmetric) Jacobi-Trudi and Giambelli 
formula (\ref{jacobi-trudi}) and (\ref{jacobi-trudi2}). 
Although we assumed that $q$ is generic, 
we expect that our NLIE (at least reduced ones 
(\ref{nlie-general})-(\ref{nlie-s=0b}), 
(\ref{nlie-xi=-1})-(\ref{nlie-r=0b})) will also be 
valid even for the case where  $q$ is root of unity 
as we will not need to take into account truncation of the 
$T$-system. 
The high temperature expansion of the free energy 
in terms of characters was calculated from our NLIE.  

There are NLIE with a finite number of unknown functions 
for algebras of arbitrary rank in different context \cite{Z98,DDT00}. 
These NLIE are different from Takahashi-type. 
Whether one can generalize (or modify) their NLIE for finite 
temperature case
 is still not clear. 
 A deeper understanding of this subject is desirable. 

There is an another kind of formulation of transfer matrices 
which is based on the graded formulation of the 
 quantum inverse scattering method.
In this formulation, the row-to-row transfer matrix 
is defined as a supertrace: 
$\widehat{t}(v)={\mathrm str}_{0}(\widehat{R}_{0L}(v)
 \cdots \widehat{R}_{02}(v)\widehat{R}_{01}(v))$, where  
the $R$-matrix is defined as $\widehat{ R}^{a_{1},b_{1}}_{a_{2},b_{2}}(v)=
(-1)^{p(a_{1})p(b_{1})}
R^{a_{1},b_{1}}_{a_{2},b_{2}}(v)$ and the graded tensor product is adopted. 
As far as the free energy (in the thermodynamic limit)
 is concerned, we think that there is no difference 
between this graded formulation and the one we have adopted. 
\section*{Acknowledgments}
The author would like to thank A. Kl\"umper and K. Sakai for 
comments on a figure of specific heats.  
He also thank Y. Nagatani for a remark 
 on programming of Mathematica. 
\noindent
\renewcommand{\theequation}{A.1.\arabic{equation}}
\begin{landscape}
\section*{Appendix: The high temperature expansion of the specific heat}
We will list the high temperature expansion of the 
specific heat $C_{sl(r+1|s+1)}$ for the $U_{q}(\widehat{sl}(r+1|s+1))$ 
Perk-Schultz model at $q=1$, 
$\mu_{a}=0$ ($a \in B$). 
Here we put $t=\frac{J}{T}$. 
In this case, $Q^{(a)}_{1}$ (cf. (\ref{limit})) becomes 
\begin{eqnarray}
Q^{(a)}_{1}=\sum_{j=0}^{a}\binom{r+1}{j}\binom{a+s-j}{a-j}, \label{Q-q=1}
\end{eqnarray}
which is the dimension of $a$-th anti-(super)symmetric tensor representation 
of $sl(r+1|s+1)$. 
If one substitute (\ref{Q-q=1}), $\Delta=1$ and the values of $(r,s)$
 into (\ref{coe1})-(\ref{coe5}), 
one can recover (\ref{hte-sl30})-(\ref{hte-sl32}) up to the order of 5 
through $C=-T\frac{\partial^{2} f}{\partial T^{2}}$. 
 A formula for $r<s$ can 
be obtained from the relation $C_{sl(s+1|r+1)}=C_{sl(r+1|s+1)}|_{t \to -t}$. 
\begin{tiny}
\begin{eqnarray}
&& \hspace{-10pt} c_{sl(3|0)}=
\frac{8 t^2}{9} + \frac{16 t^3}{27} - \frac{40 t^4}{27} - \frac{400 t^5}{243} + \frac{1246 t^6}{729} + 
   \frac{11228 t^7}{3645} - \frac{43343 t^8}{32805} - \frac{649298 t^9}{137781} + 
   \frac{120769 t^{10}}{918540} + \frac{5559367 t^{11}}{885735} + \frac{36953579 t^{12}}{18600435} - 
   \frac{4333458857 t^{13}}{584585100} - \frac{298222277909 t^{14}}{58926178080}
\nonumber \\ 
&& + 
   \frac{44130279500393 t^{15}}{5745302362800} + \frac{2885993845291237 t^{16}}{321736932316800} - 
   \frac{47755357995530701 t^{17}}{7239080977128000} - 
   \frac{4655618035381741733 t^{18}}{347475886902144000} + 
   \frac{45436230289647581 t^{19}}{12306437661117600} + 
   \frac{1590783575674338541 t^{20}}{89350942346265600} 
\nonumber \\ 
&& + 
   \frac{11365203602766081451 t^{21}}{8102505908218176000} - 
   \frac{1297481476315241042696509 t^{22}}{60606744193471956480000} - 
   \frac{8454419484929269090011049 t^{23}}{954556221047183314560000} + 
   \frac{46816780786984484371594673 t^{24}}{2016022738851651160350720} + 
   \frac{8261193033376436054715299 t^{25}}{445851182630653622000640} 
\nonumber \\ 
&& - 
   \frac{4614044757865223484648570791543 t^{26}}{208658353471145895096299520000} - 
   \frac{16687567624201209926686045552339 t^{27}}{558906303940569361865088000000} + 
   \frac{764945949570840761119882192959107 t^{28}}{45209309918748277270864896000000} + 
   \frac{4370526181353809390696401823321023 t^{29}}{104627260097674584541144473600000}
\nonumber \\ 
&& - 
   \frac{4077089856720735402715997482797183733 t^{30}}{615208289374326557101929504768000000} - 
   \frac{469645986012529902658517386886308977221 t^{31}}{8920520195927735077977977819136000000} - 
   \frac{5501623615004327359193974711230889260281 t^{32}}{583888594642542659649467639070720000000}
\nonumber \\ 
&& + 
   \frac{81517987350487140844545467182506851908591 t^{33}}{1351398263165884933985080078663680000000} + 
   \frac{8539638490748569692670776190970340550336847 t^{34}}
{273059671917403950089785323323129856000000} - 
   \frac{565870839129464697660769748292242672448332479 t^{35}}
{9097613107632737375587559089563893760000000} 
\nonumber \\ 
&&- 
   \frac{2803571976313303389947366028586799714153992385183
    t^{36}}{48253739922884039040116413411046892503040000000} + 
   \frac{821880309698434533395036032147535806012394330483
    t^{37}}{14814744713166152336877846222689835417600000000}
\nonumber \\ 
&& + 
   \frac{58773945021047530582522114436884890912882464905123
    t^{38}}{668128706624548232863150339537572357734400000000} - 
   \frac{903065874685632945085489557823621181308117028891323
    t^{39}}{24102743091480577500538148498817922805268480000000}
\nonumber \\ 
&& - 
   \frac{69053384918361487529760006169534549420582996627140401
    t^{40}}{586849397009961886969624485188610294389145600000000}
    +O(t^{41}) 
    \label{hte-sl30}
\end{eqnarray}
\begin{eqnarray}
&& \hspace{-10pt} c_{sl(2|1)}=
\frac{32 t^2}{27} + \frac{304 t^3}{729} - \frac{5480 t^4}{2187} -
 \frac{8320 t^5}{6561} + \frac{736708 t^6}{177147} + 
   \frac{1470644 t^7}{531441} - \frac{146834891 t^8}{23914845} - 
\frac{14149151840 t^9}{2711943423} + 
   \frac{228231260059 t^{10}}{27119434230} +
 \frac{3300969899909 t^{11}}{366112362105} - 
   \frac{86046211353427 t^{12}}{7908027021468}
\nonumber \\ 
&& - \frac{6841741391685967 t^{13}}{466008735193650} + 
   \frac{41669257450473325 t^{14}}{3131578700501328} +
 \frac{209509171518293955313 t^{15}}{9159867698966384400} - 
   \frac{23504662660033768183787 t^{16}}{1538857773426352579200} - 
   \frac{59568189209735825524303 t^{17}}{1731214995104646651600}
\nonumber \\ 
&& + 
   \frac{361017420632075530992718067 t^{18}}{22436546336556220604736000} + 
   \frac{3424500450806080358078749 t^{19}}{68110944235974241121520} - 
   \frac{8630572663979741453598354479 t^{20}}{588478558198817443289932800} - 
   \frac{17202731244586324123474774048501 t^{21}}{240139375283176527142516896000} 
\nonumber \\ 
&& + 
   \frac{74148256847472328975368306828013 t^{22}}{7924599384344825395703057568000} + 
   \frac{10062652817270791187839874880710933 t^{23}}{100858537618934141399857096320000} + 
   \frac{708322746602409944512187717871912169 t^{24}}{316324648705015526355199808330342400} 
\nonumber \\ 
&&- 
   \frac{15024962981052794085153555040621422973 t^{25}}{110457493727965947968161876390656000} - 
   \frac{231254246428418304396673721279263409933713 t^{26}}{9821880342290732093328954048657131520000}
 + 
   \frac{38713134042756953404786472697685030409352013 t^{27}}
{213099725283629276667762128019971692800000} 
\nonumber \\ 
&& + 
   \frac{174322192700322839649823295381972425935690253 t^{28}}{2938193181940949117691871765123852128000000} - 
   \frac{15926771354416351317222940636770636517312076257 t^{29}}
{67019011202800272994904518213769017498828800} 
\nonumber \\ 
&& - 
   \frac{668493937475506754534088912524547499759600367865761
    t^{30}}{5757542326058750725471342701091974685126656000000} 
 + 
   \frac{335250111063276140660586885786195145564952534202441623
    t^{31}}{1101993601207644888855214992989003954733241958400000}
\nonumber \\ 
&&  + 
   \frac{24356652704738251987509066739814465042906238637219292447
    t^{32}}{120217483768106715147841635598800431425444577280000000} 
- 
   \frac{238144042414431030572413060565358221701372451889917292229
    t^{33}}{626042066206424505863222816152802925415196958720000000} 
\nonumber \\ 
&&- 
   \frac{19283712635849148986736121736755878449365536043407449116181
    t^{34}}{58382894402793416775358836340649964244434367807488000000}
+O(t^{35})
\end{eqnarray}
\begin{eqnarray}
&& \hspace{-10pt} c_{sl(4|0)}=
\frac{15 t^2}{16} + \frac{15 t^3}{32} - \frac{435 t^4}{256} - \frac{345 t^5}{256} 
+ \frac{9555 t^6}{4096} + 
   \frac{21917 t^7}{8192} - \frac{172967 t^8}{65536} - \frac{3052445 t^9}{688128} 
+ \frac{53684587 t^{10}}{22020096} + 
   \frac{41381153 t^{11}}{6291456} - \frac{24190901579 t^{12}}{15854469120} 
- \frac{74629743461 t^{13}}{8304721920} - 
   \frac{59210969497 t^{14}}{186025771008} 
\nonumber \\ 
&&+ \frac{831873403828903 t^{15}}{72550050693120} + 
   \frac{40380622501051099 t^{16}}{12188408516444160} - \frac{69674366936531941 t^{17}}{5078503548518400} - 
   \frac{677763076075244557 t^{18}}{88642971028684800} 
+ \frac{27733137112330033541 t^{19}}{1808316608985169920} + 
   \frac{916368739307996439457 t^{20}}{68199369253154979840}
\nonumber \\ 
&& - 
   \frac{195301776246305171789377 t^{21}}{12368885605458562252800} - 
   \frac{71292804015129538500833 t^{22}}{3445590165496528896000} + 
   \frac{2200476451580705384154142457 t^{23}}{152384670659249486954496000} + 
   \frac{78446060458744023170123124563 t^{24}}{2681970203602790970399129600}
\nonumber \\ 
&& - 
   \frac{125238936013261236945687178727 t^{25}}{11862560515935421599842304000} - 
   \frac{191117971515476319277466700295697 t^{26}}{4934825174629135385534398464000} + 
   \frac{1065649288731276091956499544404357 t^{27}}{317238761226158703355782758400000} 
\nonumber \\ 
&&+ 
   \frac{11192351020362615444556022460039019153 t^{28}}{230949818172643536043009848115200000} + 
   \frac{156907755522484105311278338009333117 t^{29}}{19795698700512303089400844124160000} - 
   \frac{2360085804850806890473467996881990318573 t^{30}}{41081897067886708999650692982374400000} 
\nonumber \\ 
&&- 
   \frac{29423655375616416313737080966732640595067 t^{31}}{1227477288149584699201684341837004800000} + 
   \frac{3157300240979374909778816498331173586672649 t^{32}}{49099091525983387968067373673480192000000} + 
   \frac{16704208032661485316059904933586012489234849 t^{33}}
{369326254640301513906859729911545856000000} 
\nonumber \\ 
&&- 
   \frac{4632447279343820265500763653185467289421668351 t^{34}}{68884622579768236650970866196073467084800000} - 
   \frac{61387162556304338047935313287382785565501241 t^{35}}
{855075033232118492709251293240098816000000} 
\nonumber \\ 
&&+ 
   \frac{1387457285241446733857164858489037532362389241038513
    t^{36}}{21640793029659989226269007324158440419360768000000} + 
   \frac{1371535919121592468470837511909090297230678904817833
    t^{37}}{13288206246282449524902022041149919555747840000000}
\nonumber \\ 
&& - 
   \frac{20934050741656543039384025455100075026671071927370779
    t^{38}}{399522332855261339561889365984463515434352640000000}
- 
   \frac{10632927999476868411724450072247105587260120276680120833
    t^{39}}{76868096841352281731707514015410780369569447936000000}
+O(t^{40})
\end{eqnarray}
\begin{eqnarray}
&& \hspace{-10pt} c_{sl(3|1)}=
\frac{69 t^2}{64} + \frac{45 t^3}{128} - \frac{8745 t^4}{4096} -
 \frac{4065 t^5}{4096} + \frac{447405 t^6}{131072} + 
   \frac{2734683 t^7}{1310720} - \frac{406592713 t^8}{83886080}
 - \frac{222295155 t^9}{58720256} + 
   \frac{359912058803 t^{10}}{56371445760} + \frac{2122554602333 t^{11}}{338228674560} 
- 
   \frac{143332776011113 t^{12}}{18038862643200} 
\nonumber \\ 
&&- \frac{2496231619276031 t^{13}}{255121057382400} + 
   \frac{2384568827915515 t^{14}}{253987186016256} 
+ \frac{4333850790231468637 t^{15}}{297165007639019520} - 
   \frac{13960280493348579178073 t^{16}}{1331299234222807449600} - 
   \frac{6984410297833152298633 t^{17}}{332824808555701862400} + 
   \frac{633205038656776496328727 t^{18}}{58093057493358870528000} 
\nonumber \\ 
&&+ 
   \frac{14162457941778109685145689 t^{19}}{482817855611471501721600} - 
   \frac{604540593802926375008290949 t^{20}}{59593518178330196783923200} - 
   \frac{1296710847223129254977218906963 t^{21}}{32424291481573293431980032000} + 
   \frac{86141375198512658190978194262097 t^{22}}{11413350601513799288056971264000} 
\nonumber \\ 
&&+ 
   \frac{99308058853685747403471990563549 t^{23}}{1862318279967286597118853120000} - 
   \frac{49493071668124032051338572924297499 t^{24}}{22497996705704001156617901755596800} - 
   \frac{6931534160144573731055517563665688207 t^{25}}{99510370044460005115809950072832000}
\nonumber \\ 
&& - 
   \frac{587874112017235875229135097766679692569 t^{26}}{82792627876990724256353878460596224000} + 
   \frac{29312181553180719038186824685428443001969 t^{27}}{328542174115042556572832851034112000000} + 
   \frac{1698158664136245991126625182277922005467823563 t^{28}}
{77493899692863317903947230239118065664000000} 
\nonumber \\ 
&& - 
   \frac{3126686061624145346779158233961466863595139419 t^{29}}{27897803889430794445421002886082503639040000} - 
   \frac{197403265536059174349045556856630021916456067183 t^{30}}
{4463648622308927111267360461773200582246400000} 
\nonumber \\ 
&& + 
   \frac{3751792081077294678466899166499824476489567047562019
    t^{31}}{27183620109861366107618225212198791545880576000000} + 
   \frac{80680001031677274734526645657250828568042355729432743
    t^{32}}{1054394961837046927810646311261044035719004160000000}
\nonumber \\ 
&& - 
   \frac{1319799435656923687275493800075182418349748116331178151
    t^{33}}{7931220926171316229046258650147412121621626880000000} - 
   \frac{6324243289350644149071369870231550459419389543237919619
    t^{34}}{51904772136387320644826041572092537674131308544000000}
+O(t^{35})
\end{eqnarray}
\begin{eqnarray}
&& \hspace{-10pt} c_{sl(2|2)}=
\frac{9 t^2}{8} - \frac{305 t^4}{128} + \frac{4165 t^6}{1024} - \frac{1028409 t^8}{163840} + 
   \frac{758654369 t^{10}}{82575360} - \frac{775100578187 t^{12}}{59454259200} + 
   \frac{2108183654669 t^{14}}{116266106880} - \frac{36086372927030207 t^{16}}{1451001013862400} + 
   \frac{123454173470857039087 t^{18}}{3656522554933248000} 
\nonumber \\ 
&& - 
   \frac{775360975454089529227 t^{20}}{17049842313288744960} + 
   \frac{414116620493362593763666669 t^{22}}{6802887083002209239040000} - 
   \frac{18120085561666997913793728601 t^{24}}{223497516966899247533260800} + 
   \frac{4422094856669703488323197127729 t^{26}}{41123543121909461546119987200}
\nonumber \\ 
&& - 
   \frac{61544192334285763277931254839087079063 t^{28}}{433030909073706630080643465216000000} + 
   \frac{24524774786846908325218781376565239554743 t^{30}}{130948546903888884936386583881318400000} - 
   \frac{1066449426872776917417866273408875411974943 t^{32}}{4332272781704416585417709441777664000000} 
\nonumber \\ 
&&+ 
   \frac{3971063172619299668637043400605041165950069531 t^{34}}
{12300825460672899401959083249298833408000000}
+O(t^{36}) 
\label{hte-sl22}
\end{eqnarray}
\begin{eqnarray}
&& \hspace{-10pt} c_{sl(5|0)}=
\frac{24 t^2}{25} + \frac{48 t^3}{125} - \frac{1128 t^4}{625} - 
\frac{3504 t^5}{3125} + \frac{8274 t^6}{3125} + 
   \frac{178444 t^7}{78125} - \frac{1306457 t^8}{390625} - \frac{160692418 t^9}{41015625} + 
   \frac{3091451869 t^{10}}{820312500} + \frac{177674519 t^{11}}{29296875} 
- \frac{173473316029 t^{12}}{46142578125} - 
   \frac{4227122268577 t^{13}}{483398437500}
\nonumber \\ 
&& + \frac{421192483420837 t^{14}}{135351562500000} + 
   \frac{157545290437200577 t^{15}}{13196777343750000} - \frac{499112291761171129 t^{16}}{316722656250000000} - 
   \frac{20544397967491432423 t^{17}}{1319677734375000000} - 
   \frac{2522947397527190428811 t^{18}}{2217058593750000000000} 
\nonumber \\ 
&& + 
   \frac{344105756804540679754433 t^{19}}{17667185668945312500000} + 
   \frac{6481099627451266704510167 t^{20}}{1211464160156250000000000} - 
   \frac{514322825251322335204973701 t^{21}}{21971554541015625000000000} - 
   \frac{8742897574113739154665330649 t^{22}}{767260634765625000000000000}
\nonumber \\ 
&&  + 
   \frac{11410614110336974477848290940503 t^{23}}{422952424914550781250000000000} + 
   \frac{3647621524555860483885516418494301 t^{24}}{186099066962402343750000000000000} - 
   \frac{6101990866174335715593365406813719 t^{25}}{205782622121887207031250000000000} 
\nonumber \\ 
&& - 
   \frac{517943244817437855867592189832638271 t^{26}}{17121114160541015625000000000000000} + 
   \frac{21134646896102733105412988229490963037 t^{27}}{687901908236022949218750000000000000} + 
   \frac{9912970564455830832161043262688106009427 t^{28}}
{227632995089011230468750000000000000000}
\nonumber \\ 
&&  - 
   \frac{1100169921280159675483593435467841886949801 t^{29}}{37559444189686853027343750000000000000000} - 
   \frac{14414030290129165020160816003617622620734617 t^{30}}{242071098102329589843750000000000000000000} + 
   \frac{11096391762725899024492170038454077512197326053 t^{31}}
{457474030230385869873046875000000000000000000} 
\nonumber \\ 
&& + 
   \frac{1298913265483771488098839606916528515217512843843 t^{32}}
{16635419281104940722656250000000000000000000000} - 
   \frac{35827816153435004340581856161243956139756265512563
    t^{33}}{2502652047730934464599609375000000000000000000000}
\nonumber \\ 
&&  - 
   \frac{5904769218118219660609568211765655164631746178169 t^{34}}
{59827681730247539062500000000000000000000000000} - 
   \frac{93947676821931398955438041874300826546727311315759613
    t^{35}}{46799593292568474488012695312500000000000000000000000} 
\nonumber \\ 
&& + 
   \frac{6904178434557905871447348617674173162386567620702330884299
    t^{36}}{57282702190103812773327539062500000000000000000000000000}
    + 
   \frac{100656404268726322722978537972143488460344388408592120437
    t^{37}}{3823216202619438304855957031250000000000000000000000000}
\nonumber \\ 
&& - 
   \frac{286059137038561139282916743342892323995039400157684633307181
    t^{38}}{2011599909685919846554980468750000000000000000000000000000}
+O(t^{39})
\end{eqnarray}
\begin{eqnarray}
&& \hspace{-10pt} c_{sl(4|1)}=
\frac{648 t^2}{625} + \frac{912 t^3}{3125} - \frac{159176 t^4}{78125} 
- \frac{8087472 t^5}{9765625} + 
   \frac{158309102 t^6}{48828125} + \frac{10619486948 t^7}{6103515625} 
- \frac{139915063629 t^8}{30517578125} - 
   \frac{50193776068378 t^9}{16021728515625} 
+ \frac{145147029308647729 t^{10}}{24032592773437500} 
\nonumber \\ 
&&+ 
   \frac{154561110991941973 t^{11}}{30040740966796875}
 - \frac{511262973476333776477 t^{12}}{67591667175292968750} - 
   \frac{562798123639186677983 t^{13}}{70810317993164062500} + 
   \frac{27263166260140540122161 t^{14}}{3004074096679687500000} + 
   \frac{72774085086186213945892331 t^{15}}{6195902824401855468750000} 
\nonumber \\ 
&& - 
   \frac{26156239133657469488082458441 t^{16}}{2505898475646972656250000000} - 
   \frac{2149440171158051134748079150511 t^{17}}{128142535686492919921875000000} + 
   \frac{11616816820943999915517438797465393 t^{18}}{1014888882637023925781250000000000} + 
   \frac{44865031236953872796448471717341173 t^{19}}{1925570424646139144897460937500000} 
\nonumber \\ 
&& - 
   \frac{272821227014028414160064869868451715981 t^{20}}{23106845095753669738769531250000000000} 
- 
   \frac{66246387219780884484696724793606687399357 t^{21}}{2095370725728571414947509765625000000000} + 
   \frac{256119119131347681110777846055145382444833999 t^{22}}
{23049077983014285564422607421875000000000000} 
\nonumber \\ 
&& + 
   \frac{8476891798073835488652808808090088670094654917 t^{23}}{201679432351374998688697814941406250000000000} - 
   \frac{434354342487729446922438313016331279852618182267 t^{24}}
{49299416797002777457237243652343750000000000000} 
\nonumber \\ 
&& - 
   \frac{74214528639575465586270534190435386898519833666641
    t^{25}}{1352823539616565540200099349021911621093750000000} + 
   \frac{241489502492350466125582435489379440678333756529927
    t^{26}}{57851356445462442934513092041015625000000000000000}
\nonumber \\ 
&& + 
   \frac{360758243689593696611421633292891289316448278820393793389
    t^{27}}{5125268610090188303729519248008728027343750000000000000} + 
   \frac{209490702367563791064096969855392414710044548635093273111309
    t^{28}}{55967933222184856276726350188255310058593750000000000000000} 
\nonumber \\ 
&&- 
   \frac{372970955002091736988968941025597451968680727107685289984993771
    t^{29}}{4197594991663864220754476264119148254394531250000000000000000} - 
   \frac{4270970709112276953496423117872370698202654386252499239578642355083
    t^{30}}{264448484474823445907532004639506340026855468750000000000000000000} 
\nonumber \\ 
&&+ 
   \frac{705454893266763100765781948219461395410990670735178037505736499754037
    t^{31}}{6390838374808233276098690112121403217315673828125000000000000000000} + 
   \frac{40167992980468192551156832646604563149328974033305535140279044561345513
    t^{32}}{1161970613601496959290670929476618766784667968750000000000000000000000} 
\nonumber \\ 
&&- 
   \frac{137948929879317690635779621721238444356037066949039424111273171238709303
    t^{33}}{1022270328271698821254176436923444271087646484375000000000000000000000}
\nonumber \\ 
&& - 
   \frac{2953939948969279942248671840928447255547963144680752974314176599201612597751
    t^{34}}{48518201539360464871606382075697183609008789062500000000000000000000000000}
+O(t^{35})
\end{eqnarray}
\begin{eqnarray}
&& \hspace{-10pt} c_{sl(3|2)}=
\frac{672 t^2}{625} + \frac{336 t^3}{3125} - \frac{172296 t^4}{78125} 
- \frac{3045504 t^5}{9765625} + 
   \frac{179227188 t^6}{48828125} + \frac{4080719804 t^7}{6103515625}
 - \frac{168141570529 t^8}{30517578125} - 
   \frac{19837767385216 t^9}{16021728515625} 
+ \frac{1492774189466571 t^{10}}{190734863281250} + 
   \frac{63203192343466109 t^{11}}{30040740966796875} 
\nonumber \\ 
&& - \frac{484746376821625940723 t^{12}}{45061111450195312500} - 
   \frac{119657043279360047669 t^{13}}{35405158996582031250} + 
   \frac{34147743686560410475681 t^{14}}{2360343933105468750000} + 
   \frac{20015701333284035643585049 t^{15}}{3835558891296386718750000} - 
   \frac{1295881833233097980209112581867 t^{16}}{67659258842468261718750000000}
\nonumber \\ 
&&  - 
   \frac{306358982994385261087366331797 t^{17}}{39154663681983947753906250000} + 
   \frac{1412566068743727245477698757563469 t^{18}}{56382715702056884765625000000000} + 
   \frac{38642600881761570643294935210984427 t^{19}}{3369748243130743503570556640625000}
\nonumber \\ 
&&  - 
   \frac{749943826781288708352355220297511952081 t^{20}}{23106845095753669738769531250000000000} - 
   \frac{823393490024195742531909258671419921007 t^{21}}{49889779184013605117797851562500000000} + 
   \frac{30040873842462516099488089062008818909101223 t^{22}}
{720283686969196423888206481933593750000000} 
\nonumber \\ 
&& + 
   \frac{4719753433491069947419730357115496145119402591 t^{23}}
{201679432351374998688697814941406250000000000} - 
   \frac{1816573282828621256378963795157922371189821325851 t^{24}}
{34130365474848076701164245605468750000000000000} 
\nonumber \\ 
&& - 
   \frac{8322591197319584873907765768566656873942892240011 t^{25}}{253946171687857713550329208374023437500000000000} + 
   \frac{78284346134584453501416727068291711888925901387282239
    t^{26}}{1159656736020406242460012435913085937500000000000000} 
\nonumber \\ 
&& + 
   \frac{58189014600621207023093446380939189933968282293477470193
    t^{27}}{1281317152522547075932379812002182006835937500000000000} - 
   \frac{282847733617417987501884583482556260456466391798052174307
    t^{28}}{3321933358391788715380243957042694091796875000000000000}
\nonumber \\ 
&& - 
   \frac{130877056590281430510536720696452523358544877286866176351960231
    t^{29}}{2098797495831932110377238132059574127197265625000000000000000} + 
   \frac{4709049750740948620371923839229875772436358549551968355418845451943
    t^{30}}{44074747412470574317922000773251056671142578125000000000000000000}
\nonumber \\ 
&&  + 
   \frac{11079182595131990060363516585430527467174991034289759500690770327879
    t^{31}}{130425272955270066859156941063702106475830078125000000000000000000} - 
   \frac{51676596103109252978856677265778347122077560630622361028974962093289069
    t^{32}}{387323537867165653096890309825539588928222656250000000000000000000000}
\nonumber \\ 
&&  - 
   \frac{1673903716405893043635199894308218634748017092628006510068625803350703093
    t^{33}}{14567352177871708202872014226159080862998962402343750000000000000000000} 
\nonumber \\ 
&& + 
   \frac{2560112840573273875628881469331020252804964139812191437239808660979503389001
    t^{34}}{15437609580705602459147485205903649330139160156250000000000000000000000000}
+O(t^{35})
\label{hte-sl32}
\end{eqnarray}
\end{tiny}
\end{landscape}
  
\end{document}